\documentclass[prd,onecolumn,showpacs,floatfix,superscriptaddress,nofootinbib]{revtex4-2}
\usepackage{graphicx}
\usepackage{epsfig}
\usepackage{bm}
\usepackage{amssymb}
\usepackage{float}
\usepackage{amsmath}
\usepackage{dcolumn}
\usepackage{cancel}

\usepackage{mathrsfs}
\usepackage{dcolumn}
\usepackage{graphicx}
\usepackage{amsmath}
\usepackage{amsfonts}
\usepackage{amssymb}
\usepackage{microtype}
\usepackage{subfigure}
\usepackage{makeidx}
\usepackage{bm}
\usepackage{epsf}
\usepackage{color}
\usepackage{multirow,dcolumn}
\usepackage{graphicx}
\usepackage{mathrsfs}
\graphicspath{{Images/}}

\def\doi{http://doi.org}

\def\be{\begin{equation*}}
\def\ee{\end{equation*}}

\begin{document}

\title{Motion of particles in a magnetically charged Euler-Heisenberg black hole with  scalar hair}

\author{Thanasis Karakasis}
\email{thanasiskarakasis@mail.ntua.gr}
\affiliation{Physics Department, National Technical University of Athens, 15780 Zografou Campus, Athens, Greece}
	
\author{George Koutsoumbas}
	\email{kutsubas@central.ntua.gr}
	\affiliation{Physics Department, National Technical University of Athens, 15780 Zografou Campus, Athens, Greece}

\author{Eleftherios Papantonopoulos}
\email{lpapa@central.ntua.gr} \affiliation{Physics Department, National Technical University of Athens, 15780 Zografou Campus, Athens, Greece}

\vspace{4.5cm}

\begin{abstract}
We study the geodesic motion of uncharged particles in the background of a magnetically charged Euler-Heisenberg black hole with a scalar hair. The spacetime can be asymptotically (A)dS or flat and we find, analysing the behavior of the effective potential of the radial motion that in all cases there exist stable and unstable orbits. Performing numerical integrations we depict the motion of particles for planetary and critical orbits, as well as for radial geodesics. We discuss the effect that the scalar hair $\nu$, the magnetic charge $Q_{m}$ and the Euler-Heisenberg parameter $\alpha$ have on the particle motion.
\end{abstract}

\maketitle

\flushbottom

\tableofcontents

\section{Introduction}

Magnetically charged black holes have been extensively  studied mainly in connection to  their stability. In the Maxwell-Einstein theory generalized magnetically charged Reissner-Nordstr\"om black hole solutions can be found also in string theory as a generalization of the electrically charged Garfinkle-Horowitz-Strominger black hole solutions  \cite{Gibbons:1987ps,Garfinkle:1990qj}. Magnetic monopoles are closely connected to magnetically charged black holes. In \cite{Lee:1991qs} it was shown that a  magnetic monopole may be generated as a classical instability in a magnetically charged Reissner-Nordstr\"om solution. The magnetic monopoles are hypothetical particles predicted in string theories \cite{Wen:1985qj} which however have not been observed in nature. Dirac has shown that the existence of a magnetic monopole in the Universe implies the quantization of the electric charge \cite{Dirac:1931kp}.

A generalization of  Maxwell theory is the Euler-Heisenberg theory which was proposed in 1936 \cite{Heisenberg:1936nmg}. In \cite{Obukhov:2002xa} the magnetization is determined by the clouds of virtual charges surrounding the real currents and charges. A way to detect the effect of the Euler-Heisenberg theory has been proposed in \cite{Brodin:2001zz}. The coupling of the Euler-Heisenberg Lagrangian to the Ricci scalar via the volume element allowed the finding  of black holes.  One of the black hole solutions to the Euler-Heisenberg electrodynamics has been derived in \cite{Yajima:2000kw}, where analytical solutions were obtained for the magnetically and electrically charged cases and also for dyons.  Electrically charged black holes were considered in \cite{Ruffini:2013hia} and \cite{Amaro:2020xro}, and geodesics around the Euler-Heisenberg black hole has been studied in \cite{Amaro:2020xro}. Also in  \cite{Chen:2022tbb} motions of charged particles around the Euler-Heisenberg AdS black hole were studied. A study of the thermodynamics of these black holes was performed  in \cite{Magos:2020ykt, Dai:2022mko}, while the stability of these black holes, calculating the  quasinormal modes, was studied  in \cite{Breton:2021mju}. Rotating black holes were found in \cite{Breton:2019arv, Breton:2022fch}, while the Euler-Heisenberg theory in modified gravity theories in \cite{Stefanov:2007bn, Guerrero:2020uhn, Nashed:2021ctg} was studied and black hole solutions were analyzed.

A minimally coupled to gravity self interacting scalar field was introduced in the Lagrangian of the Euler-Heisenberg theory in \cite{Karakasis:2022xzm} and hairy black holes were found which they can be considered as generalization of the  Euler-Heisenberg black holes of \cite{Yajima:2000kw} and the hairy black holes \cite{Gonzalez:2013aca}. (Black holes with scalar hair in the scenario of non-linear electrodynamics have also been discussed in \cite{Barrientos:2016ubi}.) This hairy black hole solution was characterized by three parameters, the Euler-Heisenberg parameter, the magnetic charge and the scalar charge of the scalar field.
A magnetically charged hairy black hole was obtained when the Euler-Heisenberg parameter vanishes while when the scalar charge vanishes we get the black hole solution of \cite{Yajima:2000kw}. The hairy black hole solution of \cite{Gonzalez:2013aca} is recovered when both the  Euler-Heisenberg parameter and the magnetic charge vanish. It was found that scalar field dresses the black hole with a secondary scalar hair, and the size of the  black hole is getting smaller as the magnitude of the scalar field is getting larger while is getting larger as the gravitational mass is increasing. The thermodynamical properties of the found solution are interesting, as the scalar field gains entropy for the black hole by the addition of a linear term in the entropy and hence the hairy black holes are thermodynamically preferred.

The properties of  charged black holes can be revealed if one studies the geodesics around these solutions. By studying the geodesics in addition to the information we get solving the geodesic equations we also get information about the structure of the black hole. This study will allow us  to know if test particles outside the event horizon of a black hole follow stable circular orbits or not. The motion of charged particles in the Reissner-Nordstr\"om spacetime has been discussed in \cite{Olivares:2011xb}.  The geodesics of the magnetically charged Garfinkle-Horowitz-Strominger stringy black hole \cite{Garfinkle:1990qj} were analysed in \cite{Kuniyal:2015uta, Soroushfar:2016yea}, and it was found
that there exists no stable circular orbits outside the event horizon of this stringy black hole for massless test particles. In \cite{Gonzalez:2017kxt} the motion of massive particles with electric and magnetic charges in the background of a magnetically charged Garfinkle-Horowitz-Strominger stringy black hole was studied. For critical values of the magnetic charge of the black hole and the magnetic charge of the test particle   bound and unbound orbits and the two observables were found, the perihelion shift has been  studied. The trajectories depended on the electric and magnetic charges of the test particle. Furthermore, the trajectories of the hairy black holes \cite{Gonzalez:2013aca} have been discussed in \cite{Gonzalez:2015jna}, where it has been found that particles complete an oscillation in an angle less that $2\pi$. Also, the geodesics of test particles around rotating black holes were calculated in \cite{Gonzalez:2020kbv, Gonzalez:2019xfr, Gonzalez:2018lfs}.

All the possible trajectories around the Euler-Heisenberg black holes  were studied in \cite{Amaro:2020xro}. The geodesic equations were analytically integrated and the corresponding effective potentials were analysed. It was found that the stable and the unstable circular orbits of massive test particles are barely modified due to the Euler-Heisenberg non-linear contribution. For the photon trajectories it was found that the vacuum polarization effect is significant due to the  non-linear Euler-Heisenberg electromagnetic field. The influence of the angular momentum and charge of the particle around the Euler-Heisenberg AdS black hole on the Lyapunov exponent has been studied in \cite{Chen:2022tbb}. For the specific parameters of the black hole, the spatial regions where the chaos bound is violated were found by fixing the particle charge and changing its angular momentum.

In this work we will study the time-like and null geodesics in the geometry of a magnetically charged Euler-Heisenberg black hole with scalar hair \cite{Karakasis:2022xzm}. We will use the Hamilton-Jacobi formalism which is connected to the Euler-Lagrange formalism in order to obtain the equations of motion. We shall first discuss the circular orbits for asymptotically flat, dS and AdS space-times and then the null geodesics for flat space-times. We find that the scalar charge affects the size of the black hole, i.e the presence of the scalar hair shrinks the black hole and this results to a reduction of the perihelion precession. We also find that as the magnetic charge  or the scalar hair parameter is increased,  stable circular orbits appear
and this is expected because as the scalar hair parameter is increased the black hole shrinks resulting to a weak attractive force.

The work is organized as follows: In sections \ref{sec1}, \ref{sec2}, \ref{sec3} we set the basics of the paper. We introduce our model and derive the equations of motion. In section \ref{sec4} we discuss time-like geodesics, while in section \ref{sec5} we comment on the contribution of the black hole parameters to the particle motion. In section  \ref{sec6} we discuss the null geodesics, and finally in section \ref{sec7} we conclude our work and point out possible future work.

\section{The Model} \label{sec1}

We look for time-like and null geodesics in the geometry of a magnetically charged Euler-Heisenberg black hole with scalar hair \cite{Karakasis:2022xzm}.  The corresponding metric reads
\begin{equation}
    ds^{2}=-b(r)dt^{2}+\frac{dr^{2}}{b(r)}+b_{1}^{2}(r)(d\theta^{2}+\sin^{2}\theta d\phi^{2})~,
    \label{1.1}
\end{equation}
where
\begin{multline}
b(r) =c_1 r (\nu +r)+\frac{\left(2 r-c_2\right) (\nu +2 r)-4 Q_m^2}{\nu ^2}+\frac{8 \alpha  Q_m^4 \left(-\nu ^2+12 r^2+12 \nu  r\right) \left(\nu
   ^2+3 r^2+3 \nu  r\right)}{3 \nu ^6 r^2 (\nu +r)^2}+\frac{2}{\nu ^8} \ln \left(\frac{r}{\nu +r}\right)\times\\
 \left(-\nu ^5 r (c_2+\nu ) (\nu
   +r)-2 Q_m^2 r (\nu +r) \left(\nu ^4-24 \alpha  Q_m^2\right) \ln
   \left(\frac{r}{\nu +r}\right)+48 \alpha  \nu  Q_m^4 (\nu +2 r)-2 \nu ^5
   Q_m^2 (\nu +2 r)\right),\label{1.1.1}
\end{multline}
with
\begin{eqnarray}
&&b_{1}^{2}(r) = r^{2}+ r\nu \label{1.3} ~,\\
&&c_{1}= -\frac{4}{\nu^{2}}-\frac{\Lambda_{\text{eff}}}{3}~,\\
&&c_{2}=6M-\nu~.
    \label{1.1.2}
\end{eqnarray}
The scalar field that supports the hairy structure is found to be
\begin{equation} \varphi(r) = \frac{1}{\sqrt{2}}\ln\left(1+\frac{\nu}{r}\right)~,
\end{equation}
where $\nu$ plays the role of the scalar charge. This can be seen by considering an expansion of the scalar field at infinity: one may consider $z=1/r$ and expand around $z\to0$ and then substituting back $z=1/r$ one finds:
\begin{equation} \varphi(r\to\infty) \sim \frac{\nu }{\sqrt{2} r}-\frac{\nu ^2}{2 \sqrt{2} r^2}+\frac{\nu ^3}{3 \sqrt{2}
   r^3}+\mathcal{O}\left(\left(\frac{1}{r}\right)^4\right)~,\end{equation}
where $\nu$ controls the $\mathcal{O}(r^{-1})$ term in the expansion, therefore it plays the role of a scalar charge. Then, the space-time (\ref{1.1.1}) is characterized by four constants, the black hole mass $M$, the scalar charge $\nu$, the magnetic charge $Q_m$ and the Euler-Heisenberg parameter $\alpha$. We will discuss the effects of the parameters of the matter fields in the geodesic motion of neutral particles, focusing particularly on the scalar charge $\nu$.
Considering the motion of an uncharged particle, the corresponding trajectories are identified with the geodesics. Note that the vectors $K = \partial_{t}$ and $R = \partial_{\phi}$ constitute Killing vectors in our model. Therefore, the energy and the magnitude of the angular momentum of the particle are conserved
\begin{eqnarray}
&&E = - K^{\mu}p_{\mu}=\text{const.}~, \label{1.4}\\
&&J = R^{\mu}p_{\mu}=\text{const.}~. \label{1.5}
\end{eqnarray}
The Killing vectors read
\begin{eqnarray}
&&K^{\mu} = (1,0,0,0) \; \text{and} \; K_{\mu} = (-b(r),0,0,0)~,\label{1.6}\\
&&R^{\mu} = (0,0,0,1) \; \text{and} \; R_{\mu} = (0,0,0,b_{1}^{2}(r)\sin^{2}\theta)~.
\end{eqnarray}
If we consider motion in the plane $\theta = \pi/2$, we have $R_{\mu} = (0,0,0,b_{1}^{2}(r))$.
In \cite{Karakasis:2022xzm} it was shown that the examined black hole may have more than one horizons. We seek for geodesics and observables outside the event horizon. Hence, in our work, the existence of internal horizons does not make any difference. The solutions of the equation $b(r)=0$ correspond to the horizons of the black hole. For reasonable values of the parameters of the problem, such as $Q_{m} = 0.2M$, $\alpha = 0.002M^{2}$ and $\nu=1M$, we will numerically solve the above equation, in the asymptotically Anti-de Sitter, de Sitter and flat cases. In the Anti-de Sitter case $\Lambda_{\text{eff}} = -5\cdot 10^{-5}M^{-2}$, the above equation has one real positive root ($r_{H}=1.5048M$), which corresponds to the event horizon. In the de Sitter case $\Lambda_{\text{eff}} = 5\cdot10^{-5}M^{-2}$, the equation $b(r)=0$ has two real positive roots ($r_{H}=1.5050M < r_{CH}=243.4434M$), which correspond to an event horizon ($r_{H}$) and a cosmological horizon ($r_{CH}$). In the asymptotically flat case $\Lambda_{\text{eff}} = 0$, the above equation has one real positive root ($r_{H}=1.5049M$), which corresponds to the event horizon. Outside the event horizon the metric function is regular. The aforementioned chosen set of parameters are reasonable, since the Euler-Heisenberg parameter $\alpha$ constitutes a small correction to our model and the scalar hair parameter $\nu=1M$ contributes to the results without degenerating the black hole. We discuss further the contribution of the scalar hair later on.


\section{Hamilton-Jacobi formalism }\label{sec2}

In order to obtain the equations of motion, we apply the Hamilton-Jacobi formalism in our model. In this Section we show how the Hamilton-Jacobi formalism is connected to the Euler-Lagrange formalism.
From the Lagrangian formulation you can construct the Hamiltonian and then perform a canonical transformation in order to obtain the Hamilton-Jacobi equation.
In the Lagrangian formalism one defines a Lagrangian from the metric
\begin{equation}
 2 \mathcal{L} = g_{\mu \nu} \dot{x}^{\mu} \dot{x}^{\nu}~,
\end{equation}
where $\dot{x}^{\mu}=\frac{d x^{\mu}}{d \tau}$ and $\tau$ is an affine parameter, which can be the proper time for massive particles.
The Hamiltonian is given by
\begin{equation}
\mathcal{H}=p_{\mu} \dot{x}^{\mu}-\mathcal{L}=\frac{1}{2} g^{\mu \nu} p_{\mu} p_{\nu}=-\frac{m^{2}}{2}~,
\label{haml}
\end{equation}
where $d x^{\mu} / d\tau$ was expressed in terms of the conjugate momentum and the coordinates. In equation (\ref{haml}), we have considered normalised momentum ($ g_{\mu\nu} p^{\mu} p^{\nu} + m^{2} = 0$), where $m$ denotes the mass of the test particle. Equation (\ref{haml}) is valid for the case of a neutral test particle. In the case of a test particle with charge $q$, we have to substitute the momentum $p^{\mu} \rightarrow p^{\mu} - q A^{\mu}$ into equation (\ref{haml}).
Now, one can perform a canonical transformation, from the coordinates of the phase space $x^{\mu}$, $p_{\nu}$, where the Hamiltonian is $H$, to the coordinates $X^{\mu}$, $P_{\nu}$ where the Hamiltonian is $K$
\begin{equation}
p_{\mu} d x^{\mu} -H d\tau = P_{\mu} dX^{\mu} - K d\tau +dF~,
\end{equation}
where $F$ is a function of the phase space coordinates.
Considering the transformations
\begin{equation}
p_{\mu}=p_{\mu} (x^{\alpha},P_{\nu},\tau) \,\,\, , X^{\mu}=X^{\mu} (x^{\nu}, P_{\alpha}, \tau  )~,
\end{equation}
 and integrating by parts the term $P_{\mu} dX^{\mu}$, we get
 \begin{equation}
 p_{\mu} d x^{\mu} -H d\tau = - X^{\mu} dP_{\mu}- K d\tau+d (F+ X^{\mu} P_{\mu})~.\label{eq1}
 \end{equation}
 Defining the generating function $S$ of the canonical transformation as
 \begin{equation}
 S(x^{\sigma},P_{\mu}, \tau)=F+ X^{\mu} P_{\mu}~,
 \end{equation}
we obtain
 \begin{equation}
 dS=\frac{\partial S}{\partial x^{\mu}} d x^{\mu}+\frac{\partial S}{\partial P_{\mu}} dP_{\mu}+ \frac{\partial S}{\partial \tau} d \tau~.
 \end{equation}
Then equation (\ref{eq1}) gives
\begin{equation}
\left( p_{\mu}-\frac{ \partial S}{\partial x^{\mu}}  \right) d x^{\mu}-\left( \mathcal{H}+\frac{\partial S}{\partial \tau} \right) d\tau = -(X^{\mu}-\frac{\partial S}{\partial P_{\mu}}) d P_{\mu}-K d\tau~.
\end{equation}
From this equation we get
\begin{equation}
p_{\mu}=\frac{ \partial S}{\partial x^{\mu}}~,\label{2.1}
\end{equation}
\begin{equation}
X^{\mu}=\frac{\partial S}{\partial P_{\mu}}~,\label{2.2}
\end{equation}
\begin{equation}
K=\mathcal{H}(x^{\mu},p_{\sigma}, \tau)+\frac{\partial S}{\partial \tau}~.
\end{equation}
Setting the new Hamiltonian $K$ to zero we obtain a coordinate transformation for $S$
\begin{equation}
\mathcal{H}(x^{\mu},\frac{\partial S}{\partial x^{\sigma}}, \tau)+\frac{\partial S}{\partial \tau}=0~.
\end{equation}
Then using the Hamiltonian (\ref{haml}) we finally get
\begin{equation}
\frac{1}{2} g^{\mu \nu} \frac{\partial S}{\partial x^{\mu}} \frac{\partial S}{\partial x^{\nu}}+\frac{\partial S}{\partial \tau}=0~,\label{hamJac}
\end{equation}
which is the Hamilton-Jacobi equation in our case.

\section{Determination of the equations of motion}\label{sec3}

\begin{figure}
    \centering
    \includegraphics[width=0.3\textwidth]{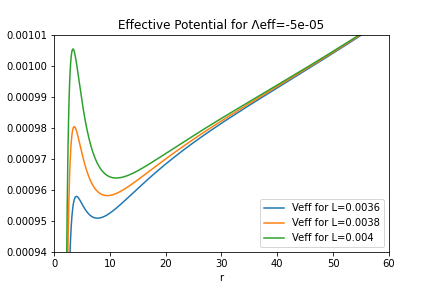}
\includegraphics[width=0.3\textwidth]{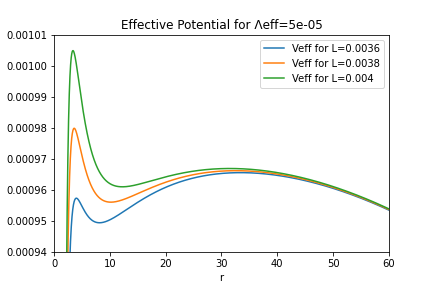}
\includegraphics[width=0.3\textwidth]{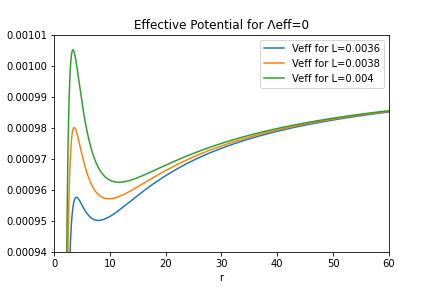}
    \caption{Effective potentials $V_{\text{eff}}$ for different values of angular momentum $L$ and $Q_{m} = 0.2M$, $m = 0.001M$, $\alpha = 0.002M^{2}$, $\nu=1M$ for asymptotically AdS (left), dS (middle) and flat (right) space-times.}
    \label{fig:2.1}
\end{figure}

Considering the metric (\ref{1.1}), the equation (\ref{hamJac}) implies

\begin{equation}
   - \dfrac{1}{b(r)} \bigg( \dfrac{\partial S}{\partial t}  \bigg)^{2} + b(r) \bigg( \dfrac{\partial S}{\partial r} \bigg)^{2} + \dfrac{1}{b_{1}^{2}(r)} \bigg( \dfrac{\partial S}{\partial \theta} \bigg)^{2} + \dfrac{1}{b_{1}^{2}(r) \sin^{2}\theta} \bigg( \dfrac{\partial S}{\partial \phi} \bigg)^{2} + m^{2} = 0~.
\label{2.8}
\end{equation}
This partial differential equation can be solved by the method of separation of variables, thus it is convenient to use the following ansatz
\begin{equation}
   S = -Et + S_{1}(r) +S_{2}(\theta) + J\phi + \frac{m^{2}}{2}\tau~,
\label{2.9}
\end{equation}
where $E,J$ can be identified as the conserved energy (\ref{1.4}) and angular momentum (\ref{1.5}) of the particle. For convenience, we substitute equation (\ref{2.1}) into the equations (\ref{1.4}) and (\ref{1.5}) and we obtain

\begin{eqnarray}
&&E = - \dfrac{\partial S}{\partial t}
\label{2.10}~,\\
&&J = \dfrac{\partial S}{\partial \phi}
\label{2.11}~.
\end{eqnarray}
Considering the ansatz (\ref{2.9}), the equation (\ref{2.8}) reads
\begin{equation}
   - E^{2} \dfrac{b_{1}^{2}(r)}{b(r)}  + b_{1}^{2}(r) b(r) \bigg( \dfrac{\partial S_{1}}{\partial r} \bigg)^{2} + \bigg( \dfrac{\partial S_{2}}{\partial \theta} \bigg)^{2} + \dfrac{J^{2}}{\sin^{2}\theta} + b_{1}^{2}(r) m^{2} = 0~.
\label{2.12}
\end{equation}
From equation (\ref{2.12}) we can identify a constant of motion
\begin{equation}
  L^{2} = \bigg( \dfrac{\partial S_{2}}{\partial \theta} \bigg)^{2} + \dfrac{J^{2}}{\sin^{2}\theta}~.
\label{2.13}
\end{equation}
In particular, without lack of generality, we may consider that the motion is developed in the invariant plane $\theta = \pi/2$. Therefore, it is obvious that $L$ is equal to angular momentum $J$. Hence, the equation (\ref{2.12}) yields
\begin{equation}
   - E^{2} \dfrac{b_{1}^{2}(r)}{b(r)}  + b_{1}^{2}(r) b(r) \bigg( \dfrac{\partial S_{1}}{\partial r} \bigg)^{2} + L^{2} + b_{1}^{2}(r) m^{2} = 0~.
\label{2.14}
\end{equation}
Thus, we can find solutions for the radial component of $S$
\begin{equation}
   S_{1}(r) = \pm \int \dfrac{dr}{b(r)} \sqrt{E^{2} - b(r) \bigg( m^{2} +\dfrac{L^{2}}{b_{1}^{2}(r)} \bigg)}~.
\label{2.15}
\end{equation}
Note that the coordinate transformation $S$ depends on $t,r,\theta,\phi$, but also it depends on $T,R,\Theta,\Phi$, which are the "new" coordinates. Due to the fact that we are looking for any coordinate transformation that satisfies the equation (\ref{hamJac}), we can set $E=E(T,R,\Theta,\Phi), J=J(T,R,\Theta,\Phi)$ as two of the ``new" coordinates. Subsequently, upon considering the equation (\ref{2.2}) and the Hamilton equations for the new coordinates, demanding that the new Hamiltonian vanishes, $K=0$, we have
\begin{equation}
   \dfrac{\partial S}{\partial E} = \text{const.}
\label{2.16}
\end{equation}
We can set the constant equal to zero, since it corresponds to the initial moment $T$. Thus, the equation (\ref{2.15}) reads
\begin{equation}
   t = \pm \int \dfrac{dr}{b(r)} E  \Bigg[ E^{2} - b(r) \bigg( m^{2} +\dfrac{L^{2}}{ b_{1}^{2}(r)} \bigg)\Bigg]^{-1/2}~.
\label{2.17}
\end{equation}
Note that we can obtain the radial velocity in the coordinate time framework
\begin{equation}
   \dfrac{dr}{dt} = \pm \dfrac{b(r)}{E} \sqrt{ E^{2} - b(r) \bigg( m^{2} +\dfrac{L^{2}}{b_{1}^{2}(r)} \bigg)}~.
\label{2.18}
\end{equation}
Considering the areas where $b(r) > 0$, we can define an effective potential:
\begin{equation}
   V_{\text{eff}}(r)=\sqrt{b(r) \bigg( m^{2} +\dfrac{L^{2}}{b_{1}^{2}(r)}\bigg)}~.
\label{2.19}
\end{equation}
For reasonable values of the parameters of our model we plot the effective potential in FIG. \ref{fig:2.1}. Note that $dr/dt=0$ at the horizons. This does not mean that nothing can cross the event horizon, instead this means that an observer outside the event horizon cannot see a particle pass through it. Upon working with the proper-time framework, we will be convinced for the analytic continuation to the inner space of the event horizon. Additionally, considering the external region to the horizon, it is obvious that the radial velocity vanishes and changes direction when $V_{\text{eff}}=E$. Consequently, it is evident that $V_{\text{eff}}$ plays the role of the effective potential of the radial motion.

\section{Time-like geodesics} \label{sec4}

In this section we will discuss the time-like geodesics of massive particles. Considering the motion of an uncharged particle with mass $m$, we will determine the time-like geodesics of the corresponding  geometry. The numerical calculation of the trajectories will be done by exploiting the Scipy integrate.quad() method using the Python programming language. This method, apart from the integral of the function, returns an estimate of the absolute error in the result, which corresponds to the integration method. The absolute errors in this paper are of the order of $10^{-8}$, and hence, do not affect the validity of the results. We will begin with the circular orbits.

\subsection{Circular orbits}

The orbits can be classified by their values of energy $E$ and angular momentum $L$. The radius $r=r_{c}$ of the circular orbits nullifies the radial derivative of the effective potential.
\begin{equation}
   \dfrac{dV_{\text{eff}}(r)}{dr} \bigg|_{r=r_{c}} = 0 \Rightarrow
   b'(r_{c})\bigg( m^{2} +\dfrac{L^{2}}{b_{1}^{2}(r_{c})}\bigg) - b(r_{c})\frac{L^{2}}{b_{1}^{4}(r_{c})}\big(b_{1}^{2}(r_{c})\big)' = 0~.
\label{2.20}
\end{equation}
Note that, according to FIG \ref{fig:2.1} in all cases, Anti-de Sitter, de Sitter and asymptotically flat, there are stable and unstable circular orbits, for our choice of parameters. Unstable circular orbits correspond to  maxima of the effective potential, while stable orbits correspond to minima of the effective potential. In the Anti-de Sitter case, for $L=0.0037 M^{2}$, $Q_{m} = 0.2M$, $m = 0.001M$, $\alpha = 0.002M^{2}$, $\nu=1M$, $M=1$ and $\Lambda_{\text{eff}} = -5\cdot10^{-5}M^{-2}$, we numerically calculate the radii of the circular orbits by solving the equation (\ref{2.20}) in the region outside the event horizons:

\begin{equation}
   \begin{matrix}
       \text{Radius\; of\;the\; unstable\; circular\; orbit:}\; r_{c}=3.788 M\\ \\
       \text{Radius\; of\;the\; stable\; circular\; orbit:}\; r_{c}=8.697 M
   \end{matrix}
\label{2.21}
\end{equation}
Similarly, we calculate the radii of the circular orbits, in the de Sitter case, for the same values of the parameters and $\Lambda_{\text{eff}} = 5\cdot10^{-5}M^{-2}$.

\begin{equation}
   \begin{matrix}
       \text{Radius\; of\;the\; inner\; unstable\; circular\; orbit:}\; r_{c}=3.779M\\ \\
       \text{Radius\; of\;the\; stable\; circular\; orbit:}\; r_{c}=9.181M\\ \\
       \text{Radius\; of\;the\; outer\; unstable\; circular\; orbit:}\; r_{c}=33.002M
   \end{matrix}
\label{2.22}
\end{equation}
Additionally, we calculate the radii of the circular orbits, in the asymptotically flat case, for the same values of the parameters and $\Lambda_{\text{eff}} =0$.

\begin{equation}
   \begin{matrix}
       \text{Radius\; of\; the\; unstable\; circular\; orbit:}\; r_{c}=3.784M\\ \\
       \text{Radius\; of\;the\; stable\; circular\; orbit:}\; r_{c}=8.919M
   \end{matrix}
\label{2.23}
\end{equation}

It is very interesting that we can generally determine the exact periods of the revolution of the circular orbits, both stable and unstable, with respect to proper time $\tau$ and coordinate time $t$. First of all, we may solve the equation (\ref{2.20}) with respect to the particle's angular momentum $L$
\begin{equation}
   L_{c} = m b_{1}^{2}(r_{c})\sqrt{\frac{b'(r_{c})}{b(r_{c})\big(b_{1}^{2}(r_{c})\big)'-b'(r_{c})b_{1}^{2}(r_{c})}}~.
\label{2.24}
\end{equation}
As we mentioned before, the equation (\ref{2.18}) implies that the condition for a vanishing radial velocity reads $E=V_{\text{eff}}(r_{c})$. Hence, the total energy of a particle, in the case of circular orbits, reads
\begin{equation}
   E_{c} = m \sqrt{b(r_{c})} \sqrt{1 +\frac{b'(r_{c})}{b(r_{c})\big(b_{1}^{2}(r_{c})\big)'-b'(r_{c})b_{1}^{2}(r_{c})}}~.
\label{2.25}
\end{equation}
The proper time period of the circular orbit is
\begin{equation}
   T_{\tau} = \frac{2\pi m b_{1}^{2}(r_{c})}{L_{c}}= 2\pi \sqrt{\frac{b(r_{c})\big(b_{1}^{2}(r_{c})\big)'-b'(r_{c})b_{1}^{2}(r_{c})}{b'(r_{c})}}~,
\label{2.26}
\end{equation}
while the coordinate time period reads
\begin{equation}
   T_{t} = T_{\tau}\frac{E_{c}}{m b(r_{c})} = \frac{2\pi b_{1}^{2}(r_{c})}{L_{c} b(r_{c})} E_{c}= \frac{T_{\tau}}{\sqrt{b(r_{c})}}\sqrt{1 +\frac{b'(r_{c})}{b(r_{c})\big(b_{1}^{2}(r_{c})\big)'-b'(r_{c})b_{1}^{2}(r_{c})}}~.
\label{2.27}
\end{equation}
It is clear that both periods are linearly dependent with the slope of their dependence being determined by the energy and mass of the particle, as well as the value of the metric function at the point of the critical orbit.
Note that the Killing vector $R=\partial_{\phi}$ defines the conservation of the magnitude of the angular momentum of the test particle
\begin{equation}
   L = R_{\mu}p^{\mu}=m b_{1}^{2}(r)\frac{d\phi}{d\tau} \Rightarrow
\label{2.28}
\end{equation}
\begin{equation}
   L_{c} =\frac{2\pi m b_{1}^{2}(r_{c})}{T_{\tau}}~.
\label{2.29}
\end{equation}
Also, the Killing vector $K=\partial_{t}$ is connected to the conservation of the total energy of the particle
\begin{equation}
   E = -K_{\mu}p^{\mu}=m b(r)\frac{dt}{d\tau} \Rightarrow
\label{2.30}
\end{equation}
\begin{equation}
   T_{t} =T_{\tau}\frac{E_{c}}{m b(r_{c})}~.
\label{2.31}
\end{equation}

\subsection{Motion in the case of a negative cosmological constant}

Having set the basics, we will now proceed to the calculation of motion of particles in the effective potential $V_{\text{eff}}$. From now on, for the calculations of the time-like geodesics, we fix the values of the parameters of the black hole and test particle in order for the observables to be straightforwardly compared. To be more precise, we consider $M=1$, $Q_{m}=0.2M$, $\alpha=0.002M^{2}$, $\nu=1M$, $m=0.001M$, while $L$ will be fixed as $L=0.0037M^{2}$ for angular motion and $L=0$ for radial motion. In the Anti-de Sitter case, we fix $\Lambda_{\text{eff}}=-5\cdot 10^{-5}M^{-2}$. All of the following figures are plotted for these values, if not stated otherwise.

\begin{figure}
    \centering
    \includegraphics[width=0.4\textwidth]{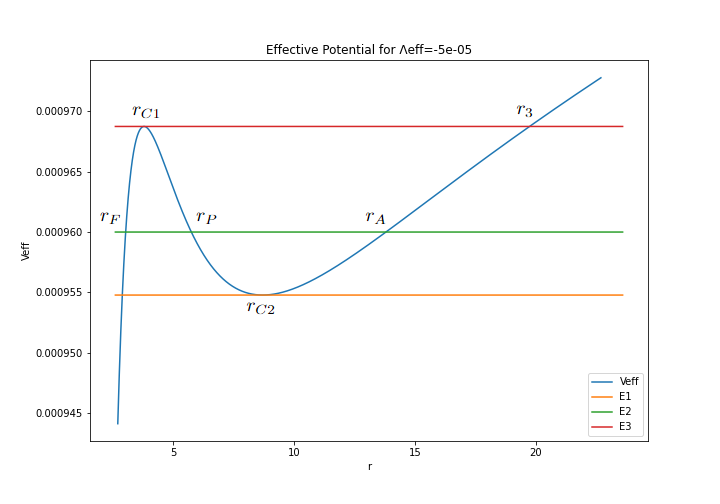}
\includegraphics[width=0.4\textwidth]{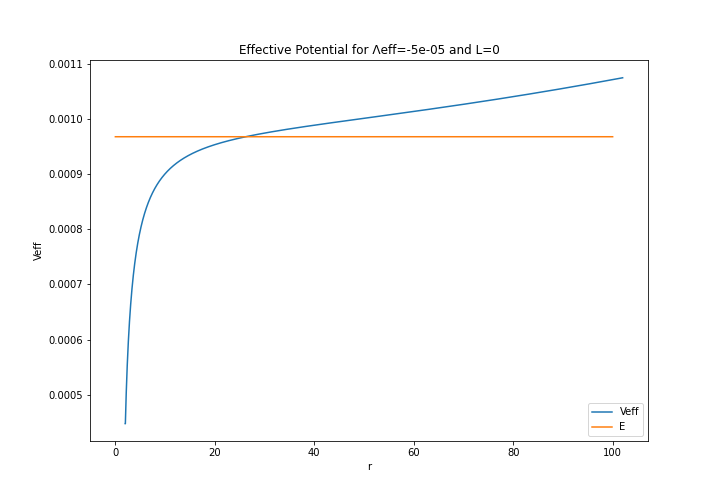}
    \caption{Effective potential for particle's motion, where $E_{1}=0.000955M$, $E_{2}=0.000960M$, $E_{3}=0.000969M$. The left figure corresponds to angular motion, while the right figure corresponds to radial motion in the AdS case.}
    \label{fig:2.4}
\end{figure}

Let us now write the equation of motion (\ref{2.18}) with respect to the proper time. The relation between the proper and the coordinate time is implied by the equation (\ref{2.30}).
\begin{eqnarray}
&&\dot{r}=\frac{dr}{d\tau}= \frac{E}{m b(r)} \frac{dr}{dt}\Rightarrow~\\
&&\dot{r}^{2}=  \frac{1}{m^{2}} \Big(E^{2} -  V_{\text{eff}}^{2}(r) \Big)~.
\label{2.32}
\end{eqnarray}
The dot indicates derivative with respect to the proper time $\tau$. The kind of motion is determined by the total energy and angular momentum of the test particle. The effective potential, for $L\neq 0$, is presented in the left panel of FIG. \ref{fig:2.4}. Considering this figure, it becomes obvious that we have stable circular motion for $E=E_{1}$ and $r=r_{C2}=8.697M$, planetary orbits for $E=E_{2}$, such that $E_{1}<E<E_{3}$, and $r_{P}\leq r\leq r_{A}$ and asymptotically unstable circular motion of radius $r=r_{C1}=3.788M$ for $E=E_{3}$. In the planetary orbits, $r_{A}$ denotes the apastron and $r_{P}$ denotes the periastron. Also, in the cases where we have $E>E_{3}$ or $E=E_{2}<E_{3}$ and $r\leq r_{F}$, the particle is doomed to fall into the event horizon. This kind of orbit will be obtained, if we choose $E=0.000956M<E_{3}$, which corresponds to $r_{F}=2.916M$. Note that the radius $r_{H}=1.505M$ of the event horizon satisfies the inequality $r_{H}<r_{F}<r_{C1}<r_{P}<r_{C2}<r_{A}<r_{3}$, where $r_{3}=19.745M$. Finally, for $L=0$, we have radial motion. The effective potential is depicted in the right panel of FIG. \ref{fig:2.4}. In this case, no stable orbits can be found and the particles are doomed to cross the event horizon and fall into the black hole.

\subsubsection{Planetary orbits}

Searching for trajectories with $L\neq 0$, we may determine the azimuthal function $\phi(r)$, since we examine the motion on the plane $\theta=\pi/2$.
\begin{eqnarray}
&&\dfrac{dr}{d\tau} = \dfrac{dr}{d\phi} \dfrac{d\phi}{d\tau} \Rightarrow \dfrac{dr}{d\phi} = \bigg(\dfrac{d\phi}{d\tau} \bigg)^{-1} \dfrac{1}{m}  \sqrt{E^{2} -  V_{\text{eff}}^{2}(r)}  \overset{(\ref{2.28})}{\Longrightarrow} \dfrac{dr}{d\phi} = \frac{b_{1}^{2}(r)}{L}  \sqrt{E^{2} -  V_{\text{eff}}^{2}(r)}\Rightarrow~,\\
&&   \phi(r)=L \int_{r_{0}}^{r}\frac{dr'}{b_{1}^{2}(r')\sqrt{E^{2} -  V_{\text{eff}}^{2}(r')}}~.
\label{2.33}
\end{eqnarray}

\begin{figure}
    \centering
    \includegraphics[width=0.4\textwidth]{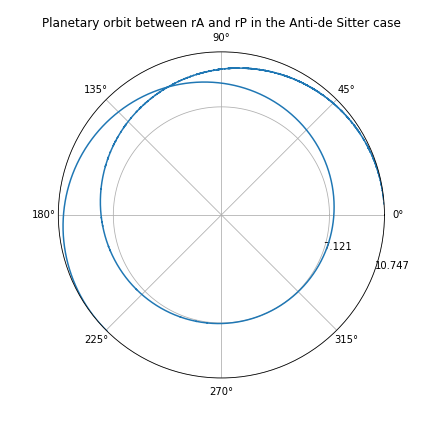}
    \caption{Planetary orbit with perihelion precession for $E=0.000956M$ in the AdS case, counterclockwise motion.}
    \label{fig:2.5}
\end{figure}

In case of planetary orbits, where we have $r_{P}\leq r\leq r_{A}$, we obtain

\begin{equation}
   \phi(r)=-L \int_{r_{A}}^{r}\frac{dr'}{b_{1}^{2}(r')\sqrt{E^{2} -  V_{\text{eff}}^{2}(r')}}~.
\label{2.34}
\end{equation}
The minus sign is a convention to obtain counterclockwise motion.
Considering a planetary orbit, a perihelion precession is observable, which basically shows the deviation of the planetary motion from a closed orbit and can be calculated by equation (\ref{2.34}) as follows
\begin{equation}
   \Delta\phi=2\phi(r_{P})-2\pi~.
\label{2.35.1}
\end{equation}
Numerical calculations of the function $\phi(r)$, for $E=0.000956M$, $r_{P}=7.121M$ and $r_{A}=10.747M$, imply a planetary orbit, with a perihelion precession, depicted in FIG.\ref{fig:2.5}. The perihelion precession reads
\begin{equation}
   \Delta\phi_{AdS}=3.963~.
\label{2.35.3}
\end{equation}

\subsubsection{Critical orbits}

\begin{figure}
    \centering
    \includegraphics[width=0.4\textwidth]{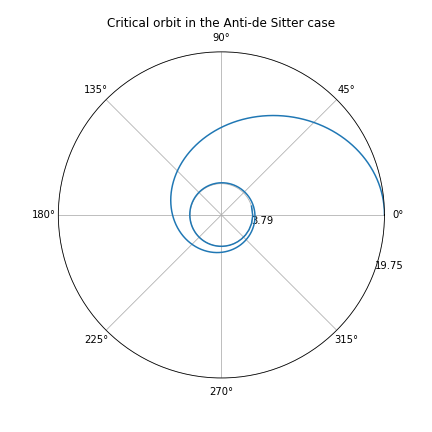}
    \caption{Critical and asymptotically circular orbit for $E=E_{3}=0.000969M$ in the AdS case. Initial position $r= r_{3}=19.75M>r_{C1}=3.79M$, counterclockwise motion.}
    \label{fig:2.6}
\end{figure}

\begin{figure}
    \centering
    \includegraphics[width=0.4\textwidth]{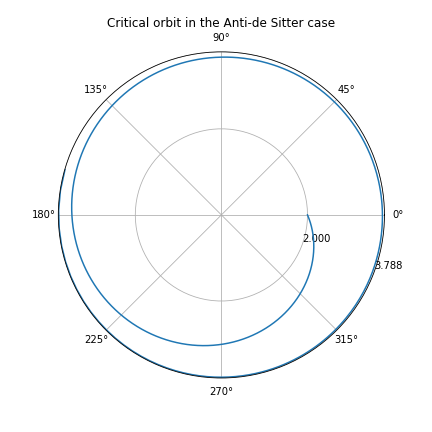}
    \caption{Critical and asymptotically circular orbit for $E=E_{3}=0.000969M$, in the AdS case. Initial position $r_{H}<r=2< r_{C1}=3.788M$, clockwise motion.}
    \label{fig:2.6.1}
\end{figure}

If the total energy of the particle takes the critical value $E_{3}=0.000969M$, we obtain asymptotically circular orbits of radius $r_{C1}=3.788M$. To be more specific, if the initial position $r$ of the particle satisfies the inequality $r_{C1}<r = r_{3}=19.745M$, where the radial velocity vanishes, the corresponding particle's critical trajectory is depicted in FIG.\ref{fig:2.6}. In this case, the particle moves from $r=r_{3}$ to $r=r_{C1}$, where the orbit becomes asymptotically circular. Additionally, upon considering an initial position $r_{H}<r< r_{C1}$, where $r_{H}=1.505M$, and a positive initial radial velocity (otherwise the particle will fall into the black hole), we obtain the trajectory depicted in FIG.\ref{fig:2.6.1}, where the particle moves from $r=2>r_{H}$ to $r=r_{C1}$, where the orbit also becomes asymptotically circular.

The period of the asymptotically circular motion approaches the values given by equations (\ref{2.26}) and (\ref{2.27}) in the proper time and coordinate time framework respectively.

\subsubsection{Falling into the event horizon}
\begin{figure}
    \centering
    \includegraphics[width=0.4\textwidth]{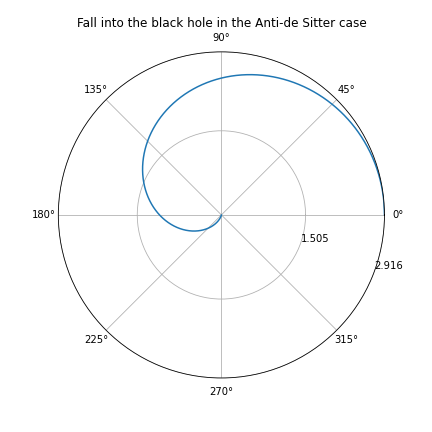}
    \caption{Orbit crossing the event horizon $r_{H}=1.505$ for $E=0.000956M$, in the AdS case. Initial position $r=r_{F}=2.916M$, counterclockwise motion.}
    \label{fig:2.7}
\end{figure}

Let us consider the motion that corresponds to $E>E_{3}$ or $E=E_{2}<E_{3}$ and $r\leq r_{F}$. In both cases the particle is doomed to fall into the event horizon. For instance, a motion with initial conditions $r=r_{F} = 2.916M$, $\frac{dr}{d\tau}=0$ and total energy $E=0.000956M$, which crosses the event horizon $r_{H}=1.505M$, is depicted in FIG. \ref{fig:2.7}. In the proper-time framework the particle falls into the black hole. Also, for our choice of parameters, the black hole has one event horizon, therefore the particle goes towards the black hole's singularity. Nevertheless, a larger value of the Euler-Heisenberg parameter may imply the existence of three event horizons and a barrier between them that may reflect the incoming particles or cause them to move in unstable circular orbits. As a result, these particles will not fall into the black-hole's singularity. In fact, three horizons imply a black hole inside of a black hole. The innermost horizon and the outermost horizon are event horizons, while the intermediate horizon is a Cauchy horizon. It is possible for a time-like particle to cross the outermost event horizon and then lie in between of the outermost and intermediate horizon, which acts as a cosmological horizon for the particle.  This phenomenon is beyond the goal of this paper and will not be investigated further. In the coordinate-time framework, the particle cannot cross the event horizon, as we will discuss in a bit.

\subsubsection{Radial trajectories}

\begin{figure}
    \centering
    \includegraphics[width=0.4\textwidth]{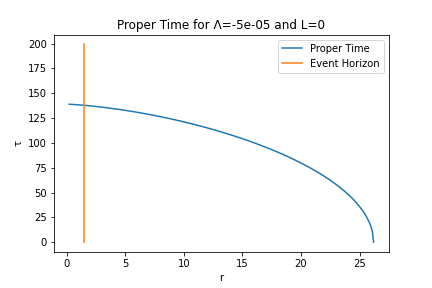}
 \includegraphics[width=0.4\textwidth]{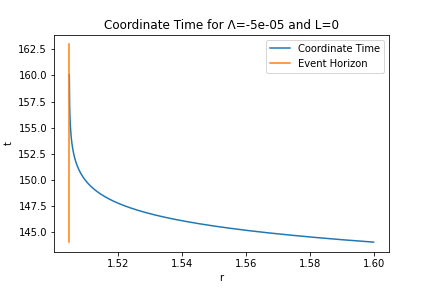}
    \caption{Radial trajectories, in the proper-time (left) and coordinate-time (right) framework, for $E=0.000968M$.}
    \label{fig:2.8}
\end{figure}

We consider the effective potential depicted in the right panel of FIG. \ref{fig:2.4} and a particle moving radially ($L=0$) with total energy $E=0.000968$. Then, equation (\ref{2.32}) yields
\begin{equation}
   \frac{dr}{d\tau}=  \frac{1}{m} \sqrt{E^{2} -  V_{\text{eff}}^{2}(r)} \Rightarrow \tau=  -m \int_{r_{F}}^{r}\frac{dr'}{\sqrt{E^{2} -  V_{\text{eff}}^{2}(r')}}~,
\label{2.36}
\end{equation}
where the initial position $r_{F}=26.201M$ corresponds to the point, where the total energy crosses the effective potential in the right panel of FIG. \ref{fig:2.4}. We present the motion that corresponds to (\ref{2.36}) in the left panel of FIG.\ref{fig:2.8}. Additionally, if we consider the equation (\ref{2.30}), we can determine the radial motion in the coordinate-time framework
\begin{equation}
   \frac{dr}{dt}=\frac{dr}{d\tau}\Big(\frac{dt}{d\tau}\Big)^{-1} \Rightarrow
   \frac{dr}{dt}= \frac{b(r)}{E}\sqrt{E^{2} -  V_{\text{eff}}^{2}(r)}  \Rightarrow
   t=  -E \int_{r_{F}}^{r}\frac{dr'}{b(r)\sqrt{E^{2} -  V_{\text{eff}}^{2}(r')}}~.
\label{2.37}
\end{equation}
The metric function $b(r)$ vanishes at the event horizon. Therefore, coordinate time, as measured by an observer outside the event horizon, must become infinite in order for the particle to cross the event horizon (right panel of FIG.\ref{fig:2.8}), while in the proper-time framework the particle falls into the event horizon in finite time (left panel of FIG.\ref{fig:2.8}).

\subsection{Motion in the case of a positive cosmological constant}

\begin{figure}[H]
    \centering
    \includegraphics[width=0.4\textwidth]{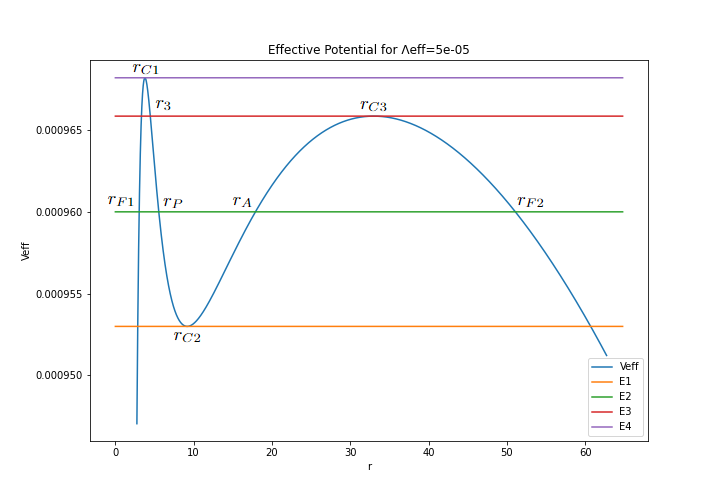}
\includegraphics[width=0.4\textwidth]{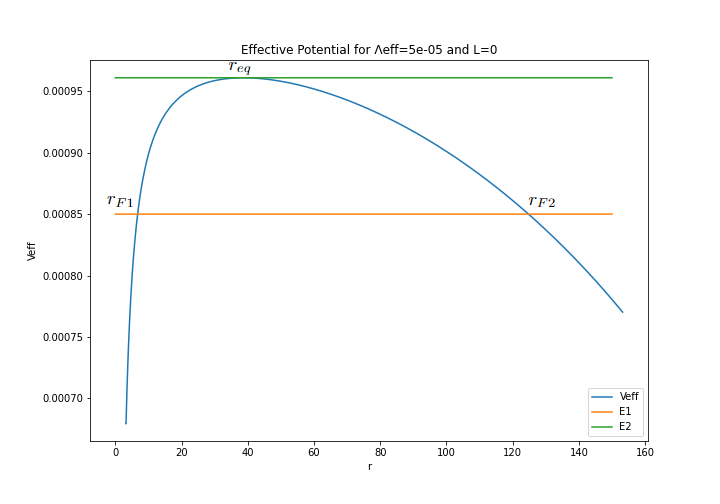}
    \caption{Effective potential of orbiting particles for $E_{1}=0.000953M$, $E_{2}=0.000960M$, $E_{3}=0.000966M$, $E_{4}=0.000968M$ (left). Effective potential of radial motion for $E_{1}=0.000850M$, $E_{2}=0.000961M$ (right). Asymptotically dS case.}
    \label{fig:2.9}
\end{figure}

We will proceed further by considering a positive cosmological constant $\Lambda_{\text{eff}} = 5\cdot 10^{-5}M^{-2}$. The kind of motion is determined by the total energy and angular momentum of the test particle, as it is obvious considering FIG. \ref{fig:2.1}. In case of angular motion, i.e. $L\neq 0$, from the left panel of FIG. \ref{fig:2.9} it follows that we have stable circular motion for $E=E_{1}$ and $r=r_{C2}=9.181M$, planetary orbits for $E_{1}<E<E_{3}$ and $r_{P}<r<r_{A}$ and asymptotically unstable circular motion of radii $r=r_{C1}=3.779M$ or $r=r_{C3}=33.002M$ for $E=E_{4}$ or $E=E_{3}$ and $r>r_{3}=4.463M$,  respectively. In the planetary orbits, $r_{A}$ denotes the apastron and $r_{P}$ denotes the periastron. Also, in the cases where we have $E<E_{4}$ and $r\leq r_{F1}$ or $r_{F2}\leq r$ the particle is doomed to fall into the event or cosmological horizon respectively. For total energy $E=0.000956M$, we obtain $r_{F1}=2.926M$ and $r_{F2}=56.974M$. The stable circular orbit for $E=E_{1}$ and $r= r_{C1}$ examined previously in section V (A. Circular orbits). Additionally, note that the radius $r_{H}=1.505M$ of the event horizon and the radius $r_{CH}=243.443M$ of the cosmological horizon satisfy the inequality $r_{H}<r_{F1}<r_{C1}<r_{3}<r_{P}<r_{C2}<r_{A}<r_{C3}<r_{F2}<r_{CH}$. Finally, for $L=0$, we have radial motion. In this case, we depict the effective potential in the right panel of FIG. \ref{fig:2.9}. The point $r_{eq}=38.637M$ corresponds to the equilibrium point between the attractive gravitational force by the black hole and the repulsive nature of the expanding universe due to the positive cosmological constant.

\subsubsection{Planetary orbits}

\begin{figure}
    \centering
    \includegraphics[width=0.4\textwidth]{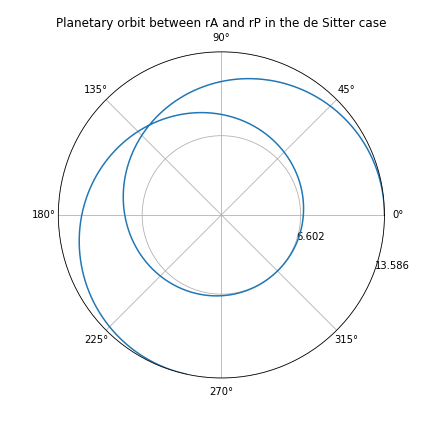}
    \caption{Planetary orbit with perihelion precession for $E=0.000956M$, in the dS case, counterclockwise motion.}
    \label{fig:3.1}
\end{figure}

We will determine the function $\phi(r)$, considering the motion on the plane $\theta=\frac{\pi}{2}$, according the procedure examined in the Anti-de Sitter case, for $L\neq 0$. In case of planetary orbits, where we have $r_{P}<r<r_{A}$, we calculate $\phi(r)$ by using equation (\ref{2.34}). The corresponding planetary orbit, depicted in FIG. \ref{fig:3.1} for $E=0.000956M$, $r_{P}=6.602M$ and $r_{A}=13.586M$, is similar to the one in the Anti-de Sitter case in FIG. \ref{fig:2.5}. This orbit has also a perihelion precession, which reads
\begin{equation}
   \Delta\phi_{dS}=4.589>\Delta\phi_{AdS}
\label{2.35.2}
\end{equation}
We can compare the perihelion precession of the Anti-de Sitter and de Sitter cases, since we use exactly the same parameters in both calculations, except for the cosmological constant, which reads $\Lambda_{\text{eff}} = -5\cdot10^{-5}M^{-2}$ and $\Lambda_{\text{eff}} = 5\cdot10^{-5}M^{-2}$ respectively.

\subsubsection{Critical orbits}

\begin{figure}
    \centering
    \includegraphics[width=0.4\textwidth]{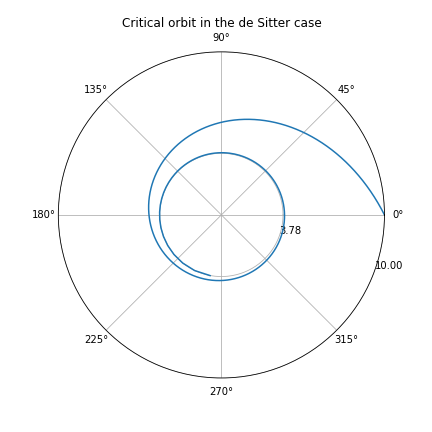}
    \caption{Critical, asymptotically circular orbit for $E=E_{4}=0.000968M$, in the dS case, with initial position $r_{C1}=3.78M<r=10M< r_{CH}$. Counterclockwise motion.}
    \label{fig:3.2}
\end{figure}

\begin{figure}
    \centering
    \includegraphics[width=0.4\textwidth]{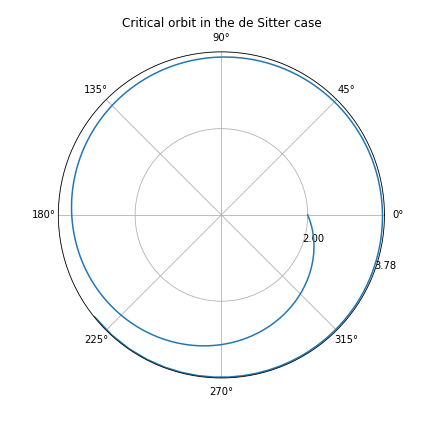}
    \caption{Critical, asymptotically circular orbit for $E=E_{4}=0.000968M$, in the dS case, with initial position $r_{H}<r=2M< r_{C1}=3.78M$. Clockwise motion.}
    \label{fig:3.2.1}
\end{figure}

\begin{figure}
    \centering
    \includegraphics[width=0.4\textwidth]{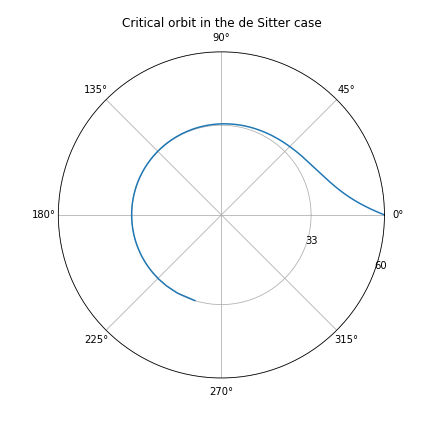}
    \caption{Critical, asymptotically circular orbit for $E=E_{3}=0.000968M$, in the dS case, with initial position $r_{C3}<r=60M< r_{CH}$. Counterclockwise motion.}
    \label{fig:3.2.2}
\end{figure}

\begin{figure}
    \centering
    \includegraphics[width=0.4\textwidth]{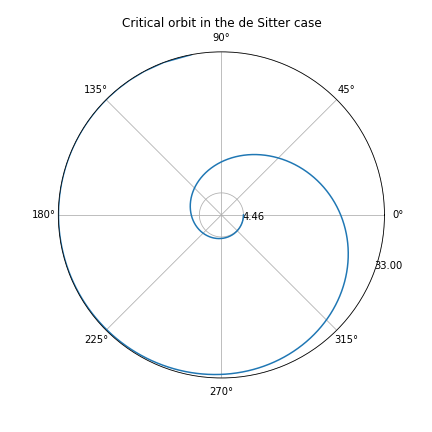}
    \caption{Critical, asymptotically circular orbit for $E=E_{3}=0.000968M$, in the dS case, with initial conditions $r= r_{3}=4.46M$, $\dot{r}=0$. Clockwise motion.}
    \label{fig:3.2.3}
\end{figure}

If the total energy of the particle takes the critical values $E_{4}=0.000968M$ or $E_{3}=0.000966M$, we obtain asymptotically circular orbits of radii $r_{C1}=3.779M$ or $r_{C3}=33.002M$, respectively. Namely, for $E=E_{4}$, if the initial position and radial velocity of the particle satisfy the inequalities $r_{C1}<r< r_{CH}$ and $\dot{r}<0$ (for initial velocity $\dot{r}>0$ the particle is doomed to cross the cosmological horizon), the corresponding particle's critical trajectory is depicted in FIG. \ref{fig:3.2}.  More specifically, the particle moves from $r=10<r_{CH}$ to $r=r_{C1}$, where the motion becomes asymptotically circular. Accordingly, upon considering different initial conditions, $r_{H}<r< r_{C1}$ and $\dot{r}>0$ (for $\dot{r}<0$ the particle will fall into the black hole), we obtain the trajectory in FIG \ref{fig:3.2.1}, where the particle moves from $r_{H}<r=2$ to $r=r_{C1}$, where the motion also becomes asymptotically circular. The case where $E=E_{3}$ is similar to the above, i.e., considering an initial position $r_{C3}<r< r_{CH}$ and a negative initial radial velocity, we obtain the trajectory presented in FIG. \ref{fig:3.2.2} (similar to FIG. \ref{fig:3.2}), while for the initial position $r= r_{3}$ we have $V_{\text{eff}}(r_{3})=E_{3}\Rightarrow\dot{r}=0$ and the corresponding trajectory is depicted in FIG. \ref{fig:3.2.3} (similar to FIG. \ref{fig:3.2.1}).

The period of the asymptotically circular motions approaches the values given by equations (\ref{2.26}) and (\ref{2.27}) in the proper time and coordinate time framework respectively.

\subsubsection{Falling into the horizons}

\begin{figure}
    \centering
    \includegraphics[width=0.4\textwidth]{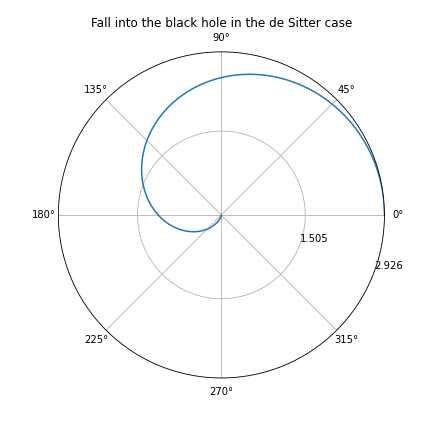}
    \caption{Orbit crossing the event horizon $r_{H}=1.505M$ for $E=0.000956M$, in the dS case, with initial position $r=r_{F1}=2.926M$. Counterclockwise motion.}
    \label{fig:3.3}
\end{figure}

Let us consider the motion that corresponds to $E>E_{4}$ and negative initial radial velocity or $E=E_{2}<E_{4}$ and $r\leq r_{F1}$. In both cases the particle is doomed to fall into the event horizon. A motion with $E=0.000956M<E_{4}$ and initial position $r=r_{F1}=2.926M$, which crosses the event horizon $r_{H}=1.505M$, is depicted in FIG. \ref{fig:3.3}. In this case, the particle's initial velocity reads $V_{\text{eff}}(r_{F1})=E\Rightarrow \dot{r}=0$. In the proper-time framework the particle falls into the event horizon and goes towards black-hole's singularity. However, in the coordinate-time framework, the particle asymptotically approaches the event horizon.

\begin{figure}
    \centering
    \includegraphics[width=0.4\textwidth]{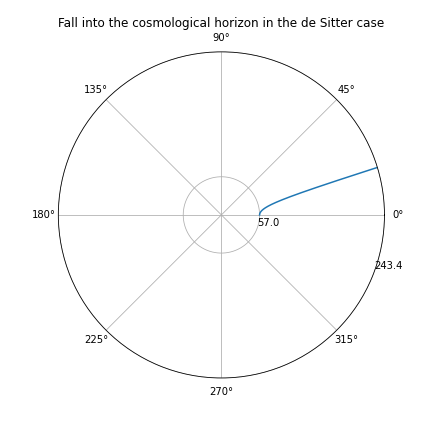}
    \caption{Orbit crossing the cosmological horizon $r_{CH}=243.443M$ for $E=0.000956M$, in the dS case, with initial position $r=r_{F2}=56.974M$.}
    \label{fig:3.4}
\end{figure}

Similarly, for $E>E_{4}$ and positive initial radial velocity or $E=E_{2}<E_{4}$ and $r\geq r_{F2}$ the particle is doomed to cross the cosmological horizon $r_{CH}=243.443M$ and lose the contact with the black hole due to the repulsive forces caused by the positive effective cosmological constant. We present the corresponding trajectory in FIG. \ref{fig:3.4}, where the particle moves from $r=r_{F2}=56.974M$ to $r=r_{CH}=243.443M$. Similarly to the case of falling into the event horizon, the particle crosses the cosmological horizon in the proper-time framework. On the other hand, in the coordinate-time framework, the particle asymptotically approaches the cosmological horizon. This may become obvious by examining the radial motion in the right panel of FIG. \ref{fig:3.5}.

\subsubsection{Radial trajectories}

Let us consider the effective potential depicted in the right panel of FIG \ref{fig:2.9} and a particle moving radially ($L=0$) with a total energy $E=0.000850M$ and an initial position $r=r_{F1}$. This case is similar to the case of the negative cosmological constant, where the analytical continuation of the space-time at the event horizon and the asymptotic motion near the event horizon, in the proper-time and coordinate-time framework respectively, is proven by calculating equations (\ref{2.36}) and (\ref{2.37}). Therefore the results are similar to those presented in FIG. \ref{fig:2.8}.

\begin{figure}
    \centering
    \includegraphics[width=0.4\textwidth]{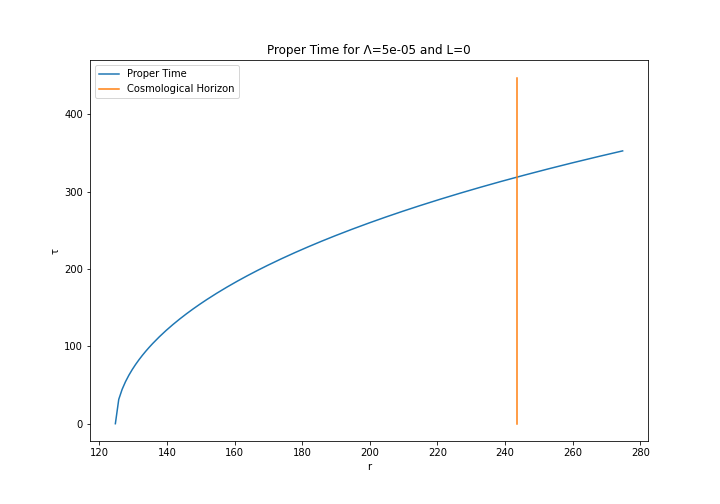}
 \includegraphics[width=0.4\textwidth]{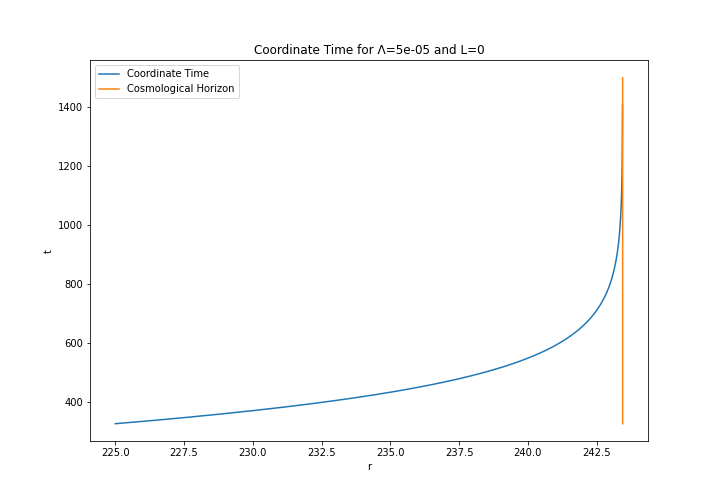}
    \caption{Radial motion in the proper-time (left) and coordinate time framework (right) for $E=0.000850M$.}
    \label{fig:3.5}
\end{figure}

Furthermore, for $E=0.000850M$, if the initial position of the particle reads $r=r_{F2}=124.770M$, the particle moves towards the cosmological horizon. We calculate the corresponding trajectories in the proper-time and coordinate-time frameworks using equations (\ref{2.36}) and (\ref{2.37}) respectively, and present the results in FIG. \ref{fig:3.5}. It is very interesting that the particle crosses the cosmological horizon in the proper-time framework, left panel of FIG \ref{fig:3.5}, while an observer, in the vicinity of the black hole, cannot see the particle falling on the cosmological horizon, right panel of FIG \ref{fig:3.5}, as in the event horizon case.

\subsection{Motion in the case of an asymptotically flat space-time}

\begin{figure}
    \centering
    \includegraphics[width=0.4\textwidth]{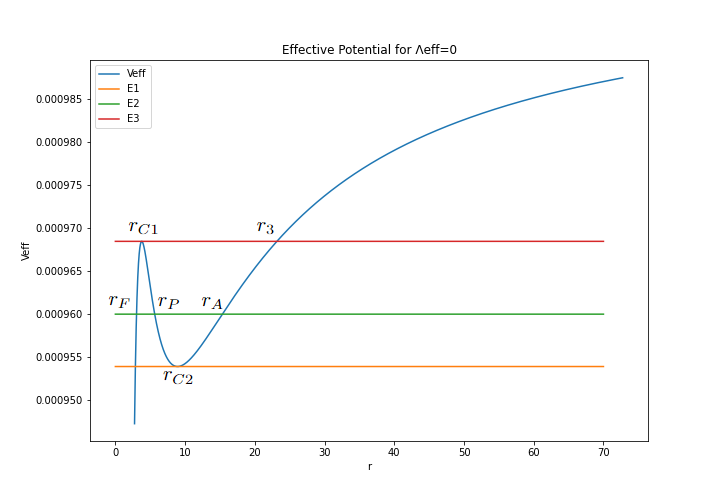}
\includegraphics[width=0.4\textwidth]{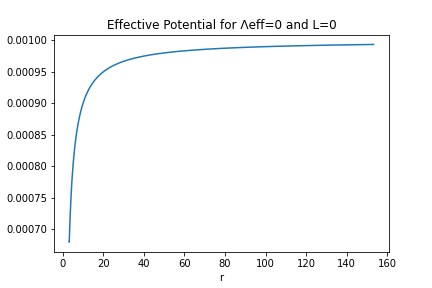}
    \caption{Effective potential of angular motion (left) for  $E_{1}=0.000954M$, $E_{2}=0.000960M$, $E_{3}=0.000968M$ and radial motion (right), in the asymptotically flat case.}
    \label{fig:4.1}
\end{figure}

The case of an asymptotically flat space-time is a special case, where we have a vanishing cosmological constant $\Lambda_{\text{eff}}=0$. The metric function asymptotically takes the value $b(r\rightarrow +\infty)=1$. In the left panel of FIG. \ref{fig:4.1} we present the effective potential for angular motion ($L\neq 0$). The effective potential for $r\rightarrow +\infty$ takes the value $V_{\text{eff}}(+\infty)=0.00100M$. Upon considering $E>0.001M$, we have motion towards the black hole or infinity, it depends on the sign of the initial radial velocity. For $E=0.001M$ and a positive initial radial velocity, we have the special case of an asymptotic motion towards infinity. For $E=E_{3}$, we have an asymptotically circular orbit of radius $r_{C1}=3.784M$. Additionally, for $E=E_{2}$ and $r_{P}<r<r_{A}$, we have a planetary orbit, also for $r\leq r_{F}$ we obtain a trajectory that crosses the event horizon and for $E=E_{1}$ and $r=r_{C2}=8.919M$ the test particle moves on a stable circular trajectory.

Moreover, considering radial motion $L=0$, the effective potential is depicted in the right panel of FIG. \ref{fig:4.1} and asymptotically reads $V_{\text{eff}}(+\infty)=0.00100M$.

\begin{figure}
    \centering
    \includegraphics[width=0.4\textwidth]{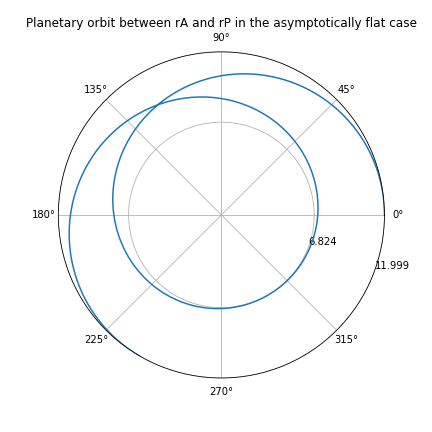}
    \caption{Planetary orbit with perihelion precession for $E=0.000956M$, in the asymptotically flat case. Counterclockwise motion.}
    \label{fig:4.3}
\end{figure}

All the above cases are similar to the (A)dS cases examined earlier. Nevertheless, we numerically calculate particle's planetary orbit presented in FIG. \ref{fig:4.3}, for $E=0.000956M$ and $r_{P}=6.824M$, $r_{A}=11.999M$. This orbit has a perihelion precession, which reads
\begin{equation}
   \Delta\phi_{fl}=4.212~.
\label{4.1}
\end{equation}
Due to the fact that we have considered the same parameters for the calculation of the perihelion precession in all cases, we present the following comparison
\begin{equation}
   \Delta\phi_{AdS}<\Delta\phi_{fl}<\Delta\phi_{dS}~.
\label{4.2}
\end{equation}
This is an expected result, considering the attractive forces imposed by the Anti-de Sitter space-time, the repulsive forces due to the positive cosmological constant in the de Sitter case and the independence of the aforementioned forces in the flat case. The equation (\ref{4.2}) yields the contribution of the effective cosmological constant to the perihelion precession, which is an observable of our model. In the next section, we discuss the contribution of the scalar hair to the perihelion precession.

\section{Contribution of the black hole parameters to the observables} \label{sec5}

It is crucial to determine the contribution of the scalar hair and magnetic charge to the trajectories of particles in our black hole spacetime. First of all, we will examine how the parameters $\nu$, $Q_{m}$ and $\alpha$ affect the effective potential. Thus, we will plot the effective potential for various values of these parameters, in the Anti-de Sitter, de Sitter and asymptotically flat cases, and compare the results. It is also important to compare the results with the $\nu=0$ case, which corresponds to a magnetically charged Euler-Heisenberg black hole without scalar hair \cite{Yajima:2000kw}, and the $Q_{m}=\alpha=0$ case, which corresponds to an asymptotically AdS black hole with scalar hair \cite{Gonzalez:2013aca}. Note here that the black hole of \cite{Gonzalez:2013aca} can be asymptotically flat or dS and not only AdS, we have checked that a curvature singularity is present and hidden behind an event horizon in each case. Furthermore, we will discuss the contribution of the scalar hair to the perihelion precession.

\subsection{Contribution of the magnetic charge to the effective potential}

\begin{figure}
    \centering
    \includegraphics[width=0.3\textwidth]{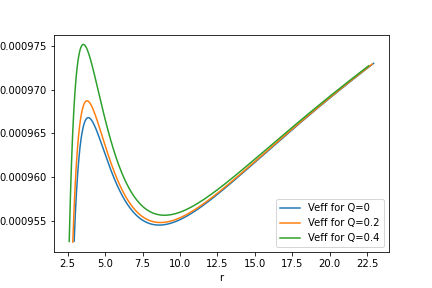}
\includegraphics[width=0.3\textwidth]{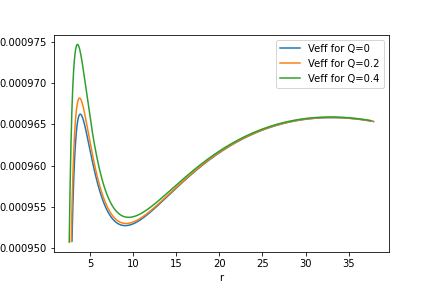}
\includegraphics[width=0.3\textwidth]{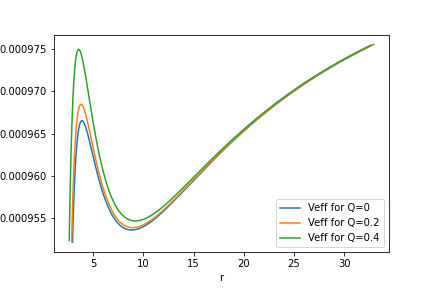}
    \caption{Effective potentials, in the AdS (left), dS (middle) and flat (right) cases for angular particle's motion for different values of the magnetic charge of the black hole $Q$.}
    \label{fig:5.1}
\end{figure}

 In order to discuss the contribution of $Q_{m}$ and $\alpha$ to the effective potential, we fix the parameters: $m = 0.001M$, $\nu=1M$ and $L=0.0037M^{2}$. It is evident that, for vanishing magnetic charge and Euler-Heisenberg parameter, our model describes an (A)dS black hole with scalar hair \cite{Gonzalez:2013aca}. In FIG. \ref{fig:5.1} we compare the effective potential in the vanishing $Q_{m}$ case with the magnetically charged Euler-Heisenberg black hole with scalar hair.

Note that, in all cases, the effective potentials are similar, i.e., there are planetary orbits and stable and unstable circular orbits in each case. However, the parameters of particle's trajectories seem to differ for each case. For instance, the radii of the stable and unstable circular orbits and the periastron and apastron of the planetary orbits are some of the observable differences. In addition, we invite the reader to note that a larger magnetic charge, results to a larger peak for the effective potential and the repulsive gravitational nature of the magnetic charge will result to a black hole space-time, in which the particles will need higher energy (keeping the angular momentum fixed) to fall in the black hole, just like the Reissner-Nordstrom case.

\begin{figure}[h]
    \centering            \includegraphics[width=0.4\textwidth]{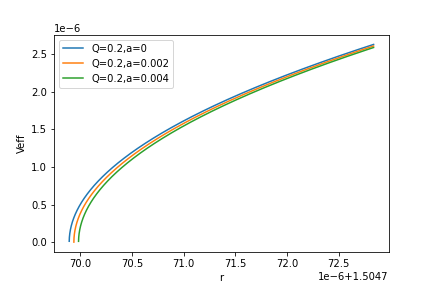}
    \caption{Effective potential, in the Anti-de Sitter case, for angular particle's motion near the event horizon for different values of $\alpha$ and $Q_{m}=0.2M$, $m = 0.001M$, $\nu=1M$, $L=0.0037M^{2}$ and $\Lambda_{\text{eff}} = -5\cdot10^{-5}M^{-2}$. The radial coordinate $r$ is depicted for values of the form $1.5047+()\cdot10^{-6}M = r_H+ \Delta r_H$ where $\Delta r_H$ is of order $10^{-6}M$.}
    \label{fig:5.3.1}
\end{figure}

Moreover, we cannot help but mention that the Euler-Heisenberg correction does not affect the effective potential far away from the event horizon. In fact, at large distances the effect of the Euler-Heisenberg parameter appears as a $\mathcal{O}(r^{-6})$ term, meaning that it is completely negligible
\begin{equation}
    b(r\to\infty) \sim -\frac{2\alpha Q_{m}^{4}}{5r^{6}}+\mathcal{O}\left(\frac{1}{r}\right)^{7}~.
    \label{5.1}
\end{equation}
 However, the Euler-Heisenberg parameter may determine the number of the horizons of the black hole, as we mentioned earlier, and contribute to the effective potential near the event horizon, as we depict in FIG. \ref{fig:5.3.1}. In particular, in FIG. \ref{fig:5.3.1}, we plot the effective potential for three different values of $\alpha$. The plots start from the event horizon of each case. Note that, as we increase the value of $\alpha$, the event horizon becomes slightly bigger. Also, it is obvious that, for $r>r_{H}+10^{-6}$, the three cases are indistinguishable. In conclusion, the Euler-Heisenberg parameter $\alpha$ does not play a significant role in the trajectories of uncharged particles, however in general it is expected that $\alpha$ will play a dramatic role near the singularity of the black hole. Note that in our model $\alpha$ controls the behaviour of space-time near the origin (equations (18) and (26) in  \cite{Karakasis:2022xzm}).

\subsection{Contribution of the scalar hair to the effective potential and perihelion precession}

\begin{figure}[h]
    \centering            \includegraphics[width=0.4\textwidth]{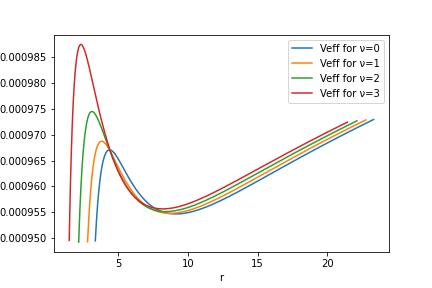}
\includegraphics[width=0.4\textwidth]{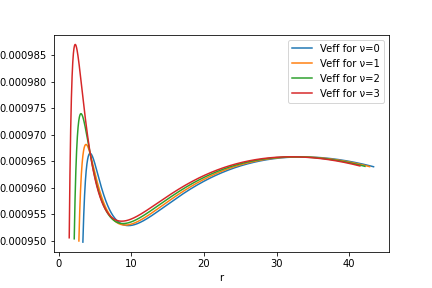}
\includegraphics[width=0.4\textwidth]{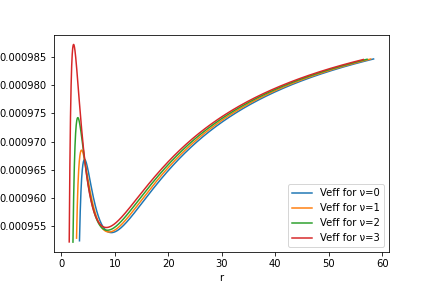}
\includegraphics[width=0.35\textwidth]{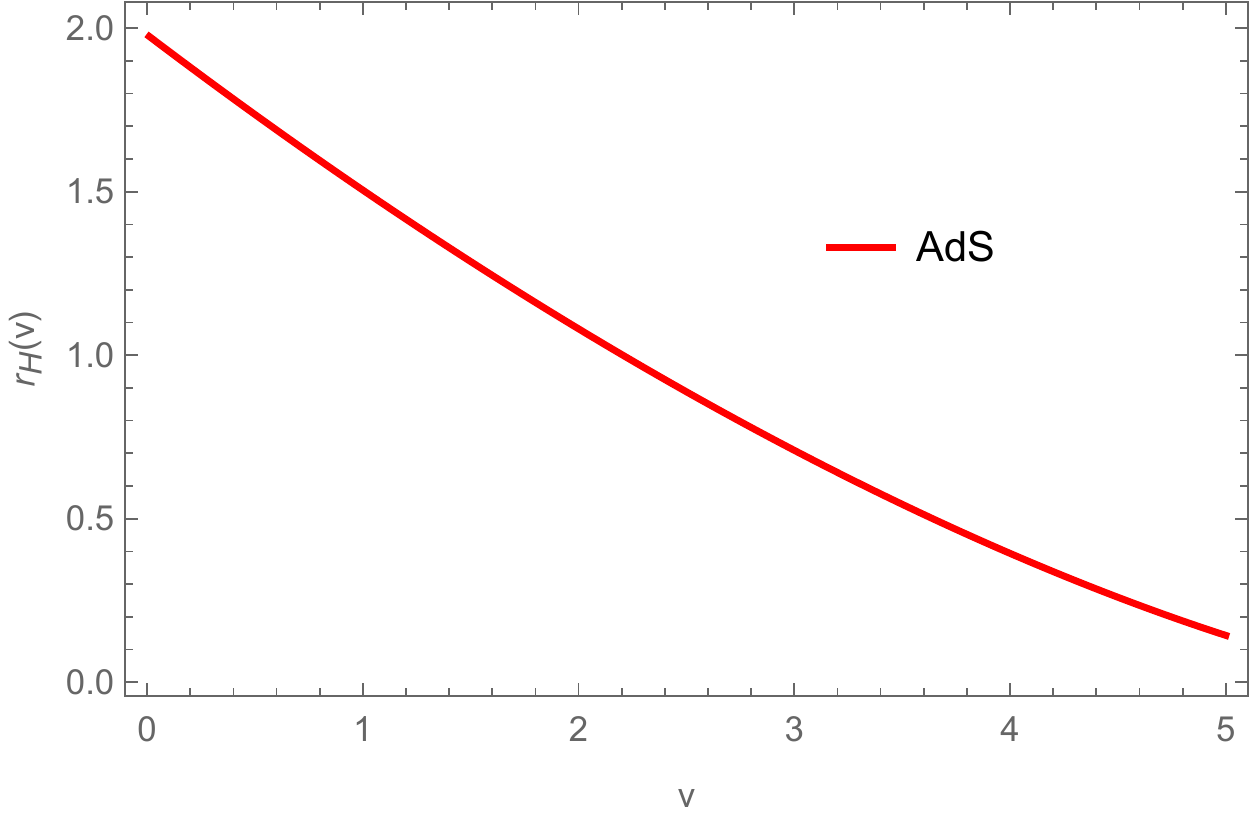}
    \caption{Effective potential, in the AdS (upper left), dS (upper right) and flat (down left) scenario for different values of $\nu$. We also plot the event horizon radius (bottom right) while varying the scalar charge $\nu$.}
    \label{fig:5.7}
\end{figure}

\begin{figure}[h]
    \centering
    \includegraphics[width=0.4\textwidth]{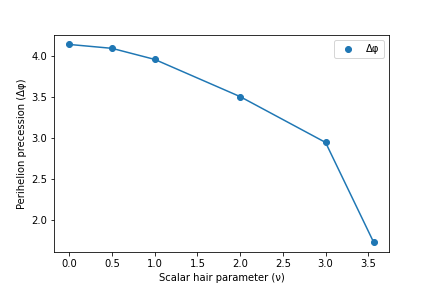}
    \caption{Perihelion precession of particle's planetary orbits, in the AdS case, for different values of $\nu$ having fixed $E=0.000956M$, $m = 0.001M$, $Q_{m}=0.2M$, $\alpha=0.002M^{2}$, $L=0.0037M^{2}$ and $\Lambda_{\text{eff}} = -5\cdot10^{-5}M^{-2}$.}
    \label{fig:5.12}
\end{figure}

In this subsection, in order for the results to be straightforwardly compared, we fix the values of the parameters $m = 0.001M$, $Q_{m}=0.2M$, $\alpha=0.002M^{2}$ and $L=0.0037M^{2}$, before proceeding to the following discussion. The scalar hair barely contributes to the effective potential for $\nu<0.1M$. In fact, the corrections of the scalar hair become significant for $\nu\geq 1M$, that is why we determined the time-like geodesics for $\nu=1M$ earlier. For convenience, we present the effective potential in the case of a magnetically charged black hole without scalar hair ($\nu=0$) compared with the case of the black hole with scalar hair, for several values of $\nu$, in FIG. \ref{fig:5.7}. For these values of $\nu$ parameter, the scalar hair plays an important role in our model by affecting the observables of the model rather than the different types of geodesics. For a fixed value of the particle's angular momentum $L$, the potential is getting larger as we increase the scalar hair parameter, meaning that the black hole shrinks in size and it is more difficult for a particle to fall into the black hole, since as we increase $\nu$ the particle's energy $E$ needs to be increased. The behaviour of the scalar hair parameter has the same effect as the magnetic charge. Both parameters have the opposite gravitational effect when compared to the mass of the black hole (a large gravitational mass gives a large black hole). We also plot in FIG. \ref{fig:5.7} the radius of the event horizon as a function of the scalar charge $\nu$ i.e $b(r_H)=0 \to r_H = r_H(\nu)$ in the AdS case, where it is clear that the black hole shrinks. We have checked that this is also true in the dS and flat cases.
Additionally, it is essential to test how the scalar hair parameter contributes to the perihelion precession. Consequently, we present the perihelion precession, $\Delta\phi=2\phi(r_{P})-2\pi$, with respect to $\nu$, in FIG. \ref{fig:5.12}, where it is clear that the perihelion precession is reduced, and the orbits tend to be more ``closed".

\section{Null geodesics}\label{sec6}

\begin{figure}
    \centering
    \includegraphics[width=0.4\textwidth]{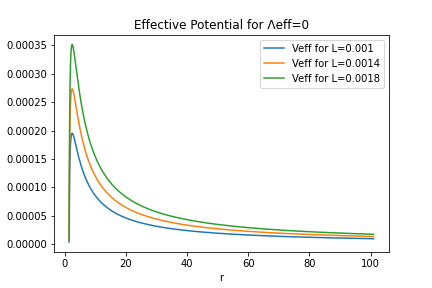}
\includegraphics[width=0.4\textwidth]{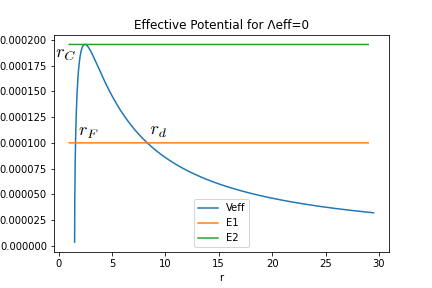}
    \caption{Effective potential of photon's angular motion, in the asymptotically flat case, for different values of angular momentum $L$ (left) and a particular case of the effective potential for $L=0.001M^{2}$, $E_{1}=0.000100M$, $E_{2}=0.000195M$ (right).}
    \label{fig:6.1}
\end{figure}

In this section, we will consider the motion of an uncharged massless particle and we will determine the null geodesics of the examined geometry. Considering $m=0$ in the previous analysis about the time-like geodesics, the equations (\ref{2.18}) and (\ref{2.19}) yield
\begin{equation}
 \dfrac{dr}{dt} = \pm \dfrac{b(r)}{E} \sqrt{ E^{2} -  \dfrac{b(r)}{b_{1}^{2}(r)}L^{2}}~,
\label{6.1}
\end{equation}
and
\begin{equation}
   V_{\text{eff}}(r)=L\sqrt{ \dfrac{b(r)}{b_{1}^{2}(r)}}~.
\label{6.2}
\end{equation}
The effective potential is defined in the regions where $b(r)\geq 0$. The three considered cases, Anti-de Sitter, de Sitter and asymptotically flat, do not have any serious differences, except for the de Sitter case that includes a cosmological horizon. Thus, we will proceed further, considering the asymptotically flat case, in order to determine the geodesics in the vicinity of the black hole. For the following discussion, we fix the parameters $Q_{m}=0.2M$, $\alpha=0.002M^{2}$, $\nu=1$ and $\Lambda_{\text{eff}} =0$. We depict the effective potential for the angular motion, in the left panel of FIG. \ref{fig:6.1}, for different values of photon's angular momentum $L$. Then, we fix the value of $L$ and investigate the different types of motion that correspond to the different values of the total energy of the photon, right panel of FIG. \ref{fig:6.1}. In particular, for $E>E_{2}$ or $E=E_{1}<E_{2}$ and $r\leq r_{F}$ the photon falls into the black hole. Also, for $E<E_{2}$ and $r>r_{d}$, the photon deflects on the effective potential. An interesting case corresponds to the critical value of the energy $E=E_{2}$, since, in this case, the photon follows an asymptotically circular trajectory of radius $r_{C}$.

\begin{figure}[h]
    \centering
    \includegraphics[width=0.4\textwidth]{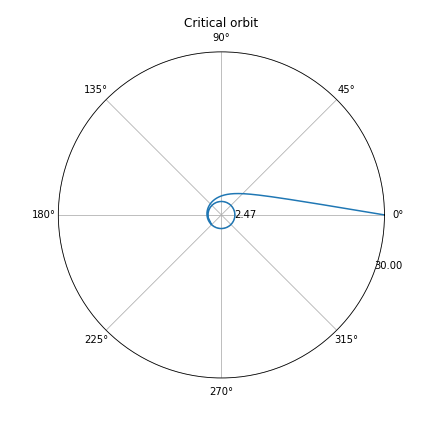}
\includegraphics[width=0.4\textwidth]{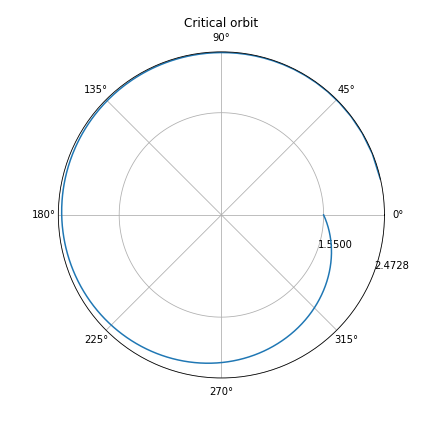}
    \caption{Asymptotically circular orbits, for  $E=E_{2}=0.000195M$. In the left panel we have particles that come from $r=30M>r_{C}=2.4728M$ and have a negative radial velocity and in the right panel we have particles that come from $r: r_H<r=1.55M<r_C$ and have a positive initial radial velocity.}
    \label{fig:6.3}
\end{figure}

The asymptotically circular orbits are classified into two categories. In the first category, left panel of FIG. \ref{fig:6.3}, we have orbits with initial conditions: $r>r_{C}=2.473M$ and negative radial velocity, while, in the second category, right panel of FIG. \ref{fig:6.3}, we have orbits with initial conditions: $r_{H}<r<r_{C}$ and positive radial velocity. For the chosen set of parameters, the radius of the event horizon is $r_{H}=1.505M$.

\begin{figure}[h]
    \centering
    \includegraphics[width=0.4\textwidth]{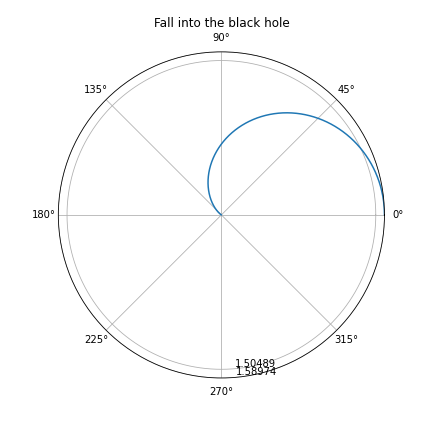}
    \caption{The photon falls into the black hole, $E=0.000100M$. Initial position $r:$ $r=r_{F}=1.58974M>r_{H}=1.50489M$. Counterclockwise motion.}
    \label{fig:6.5}
\end{figure}

Furthermore, for initial position $r=r_{F}=1.5897M$ and total energy $E=E_{1}=0.000100M$, the photon is doomed to cross the event horizon and fall into the singularity, FIG. \ref{fig:6.5}.

\begin{figure}
    \centering
    \includegraphics[width=0.4\textwidth]{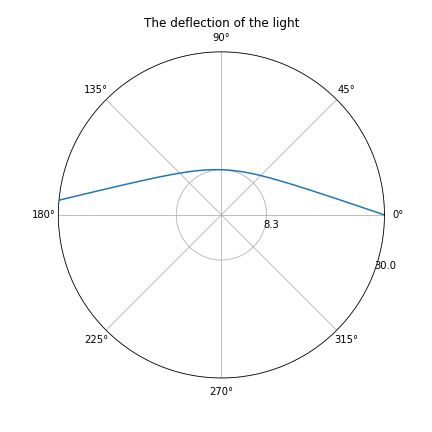}
    \caption{The deflection of light on the black hole, for $L=0.001M^{2}$, $E=0.000100M$, $Q_{m}=0.2M$, $\alpha=0.002M^{2}$, $\nu=1$ and $\Lambda_{eff} =0$. Initial position $r=30M>r_{d}=8.3M$. Counterclockwise motion.}
    \label{fig:6.6}
\end{figure}

Additionally, the deflection of light on the black hole is an interesting phenomenon that corresponds to the curved space-time around the black hole. Similarly to the angular motion in the massive case, we determine the azimuthal angle

\begin{equation}
   \phi(r)=-L \int_{r_{0}}^{r}\frac{dr'}{b_{1}^{2}(r')\sqrt{E^{2} -  V_{\text{eff}}^{2}(r')}}~,
\label{6.2.1}
\end{equation}
where $r_{0}>r_{d}$ is an arbitrary initial position. We assume motion on the $\theta=\pi/2$ plane. In FIG. \ref{fig:6.6}, we depict the corresponding trajectory for total energy $E=E_{1}=0.000100M$ and $r_{d}=8.3M$. To be more precise, the photon moves from $r=30M>r_{d}$ to $r=r_{d}$, where it changes direction (deflection of light) and starts moving away from the black hole.
Finally, considering radial motion ($L=0$), the effective potential (\ref{6.2}) vanishes. Hence the radial velocity, in the coordinate-time framework, reads
\begin{equation}
   \dfrac{dr}{dt} = \pm b(r)~.
\label{6.3}
\end{equation}
In such a case, it is evident that a regular observer, located outside of the event horizon of the black hole, testifies that light cannot cross the event horizon or the cosmological horizon in the de Sitter case. Also, as it is usual, in the proper-time framework, the radius and the proper-time are linearly dependent. Considering that $\frac{dt}{d\tau}=\frac{E}{b(r)}$, the equation (\ref{6.3}) yields
\begin{equation}
   \dfrac{dr}{d\tau} = \pm E=\text{const.}
\label{6.4}
\end{equation}
Consequently, the continuation  of the space-time, inside and outside of the event horizon, becomes obvious, and hence, the photon in the proper-time framework crosses the horizons. We do not present the corresponding figures, since they are similar to the cases examined in the previous sections.

\section{Conclusions}

\label{sec7}

In this work, we examined the trajectories of uncharged particles, in a magnetically charged black hole space-time dressed with a scalar hair in the theory of Euler-Heisenberg electrodynamics. Both time-like and null cases have been considered. We found that the magnetic charge affects the particle motion in a repulsive way, in the sense that for a fixed angular momentum, particles require higher energy in order to move in this black hole space-time. The scalar charge affects the motion in the same manner, while the modified electrodynamics Euler-Heisenberg parameter has minimal effects on the particle motion. However, we have concluded that the Euler-Heisenberg parameter will play a decisive role near the black hole singularity. It also plays a significant role in the causal structure of space-time, since it might give black holes with three horizons, two event and one intermediate (Cauchy) horizon, introducing in this way the notion of a black hole inside of a black hole.

We have numerically integrated the radial equation of motion and treating the $r,\phi$ coordinates as polar coordinates we have plotted the different kinds of motion, in the AdS, dS and flat cases. We found that the perihelion shift is getting smaller when the scalar charge grows stronger and that the dS case has a larger perihelion shift, because of the repulsive nature of a positive cosmological constant.

It would be interesting to consider the motion of charged particles in this space-time, in order to see what the effect of the Euler-Heisenberg parameter will be in this case. The geodesic motion of particles in the background of space-times with a magnetic monopole induced from a global monopole with a well-defined AdM mass such as \cite{Chatzifotis:2022ubq} may be considered in a future work.
The shadow cast by these black hole space-times would also be worth to be considered, in a similar manner, as in \cite{Khodadi:2020jij}.


\begin{thebibliography}{99}

\bibitem{Gibbons:1987ps}
  G.~W.~Gibbons and K.~i.~Maeda,
  ``Black Holes and Membranes in Higher Dimensional Theories with Dilaton Fields,''
  Nucl.\ Phys.\ B {\bf 298}, 741 (1988).

\bibitem{Garfinkle:1990qj}
  D.~Garfinkle, G.~T.~Horowitz and A.~Strominger,
  ``Charged black holes in string theory,''
  Phys.\ Rev.\ D {\bf 43}, 3140 (1991)
  Erratum: [Phys.\ Rev.\ D {\bf 45}, 3888 (1992)].

\bibitem{Lee:1991qs}
  K.~M.~Lee, V.~P.~Nair and E.~J.~Weinberg,
  ``A Classical instability of Reissner-Nordstrom solutions and the fate of magnetically charged black holes,''
  Phys.\ Rev.\ Lett.\  {\bf 68}, 1100 (1992)
  [hep-th/9111045].


\bibitem{Wen:1985qj}
  X.~G.~Wen and E.~Witten,
  ``Electric and Magnetic Charges in Superstring Models,''
  Nucl.\ Phys.\ B {\bf 261}, 651 (1985).

   \bibitem{Dirac:1931kp}
  P.~A.~M.~Dirac,
  ``Quantized Singularities in the Electromagnetic Field,''
  Proc.\ Roy.\ Soc.\ Lond.\ A {\bf 133}, 60 (1931).



\bibitem{Barriola:1989hx}
  M.~Barriola and A.~Vilenkin,
  ``Gravitational Field of a Global Monopole,''
  Phys.\ Rev.\ Lett.\  {\bf 63}, 341 (1989).


\bibitem{Harari:1990cz}
D.~Harari and C.~Lousto,
``Repulsive gravitational effects of global monopoles,''
Phys. Rev. D \textbf{42} (1990), 2626-2631
doi:10.1103/PhysRevD.42.2626


\bibitem{Heisenberg:1936nmg}
W.~Heisenberg and H.~Euler,
``Consequences of Dirac's theory of positrons,''
Z. Phys. \textbf{98} (1936) no.11-12, 714-732
[arXiv:physics/0605038 [physics]].

\bibitem{Obukhov:2002xa}
Y.~N.~Obukhov and G.~F.~Rubilar,
``Fresnel analysis of the wave propagation in nonlinear electrodynamics,''
Phys. Rev. D \textbf{66} (2002), 024042
[arXiv:gr-qc/0204028 [gr-qc]].

\bibitem{Brodin:2001zz}
G.~Brodin, M.~Marklund and L.~Stenflo,
``Proposal for Detection of QED Vacuum Nonlinearities in Maxwell's Equations by the Use of Waveguides,''
Phys. Rev. Lett. \textbf{87} (2001), 171801
[arXiv:physics/0108022 [physics.class-ph]].

\bibitem{Yajima:2000kw}
H.~Yajima and T.~Tamaki,
``Black hole solutions in Euler-Heisenberg theory,''
Phys. Rev. D \textbf{63} (2001), 064007
[arXiv:gr-qc/0005016 [gr-qc]].

\bibitem{Ruffini:2013hia}
R.~Ruffini, Y.~B.~Wu and S.~S.~Xue,
``Einstein-Euler-Heisenberg Theory and charged black holes,''
Phys. Rev. D \textbf{88} (2013), 085004
[arXiv:1307.4951 [hep-th]].

\bibitem{Amaro:2020xro}
D.~Amaro and A.~Mac\'\i{}as,
``Geodesic structure of the Euler-Heisenberg static black hole,''
Phys. Rev. D \textbf{102} (2020) no.10, 104054

\bibitem{Chen:2022tbb}
D.~Chen and C.~Gao,
``Circular motion and chaos bound of charged particles around Einstein-Euler-Heisenberg AdS black hole,''
[arXiv:2205.08337 [hep-th]].

\bibitem{Magos:2020ykt}
D.~Magos and N.~Bret\'on,
``Thermodynamics of the Euler-Heisenberg-AdS black hole,''
Phys. Rev. D \textbf{102} (2020) no.8, 084011
[arXiv:2009.05904 [gr-qc]].

\bibitem{Dai:2022mko}
H.~Dai, Z.~Zhao and S.~Zhang,
``Thermodynamic phase transition of Euler-Heisenberg-AdS black hole on free energy landscape,''
[arXiv:2202.14007 [gr-qc]].

\bibitem{Breton:2021mju}
N.~Bret\'on and L.~A.~L\'opez,
``Birefringence and quasinormal modes of the Einstein-Euler-Heisenberg black hole,''
Phys. Rev. D \textbf{104} (2021) no.2, 024064
[arXiv:2105.12283 [gr-qc]].


\bibitem{Breton:2019arv}
N.~Bret\'on, C.~L\"ammerzahl and A.~Mac\'\i{}as,
``Rotating black holes in the Einstein\textendash{}Euler\textendash{}Heisenberg theory,''
Class. Quant. Grav. \textbf{36} (2019) no.23, 235022

\bibitem{Breton:2022fch}
N.~Bret\'on, C.~L\"ammerzahl and A.~Mac\'\i{}as,
``Rotating structure of the Euler-Heisenberg black hole,''
Phys. Rev. D \textbf{105} (2022) no.10, 104046

\bibitem{Stefanov:2007bn}
I.~Z.~Stefanov, S.~S.~Yazadjiev and M.~D.~Todorov,
``Scalar-tensor black holes coupled to Euler-Heisenberg nonlinear electrodynamics,''
Mod. Phys. Lett. A \textbf{22} (2007), 1217-1231
[arXiv:0708.3203 [gr-qc]].

\bibitem{Guerrero:2020uhn}
M.~Guerrero and D.~Rubiera-Garcia,
``Nonsingular black holes in nonlinear gravity coupled to Euler-Heisenberg electrodynamics,''
Phys. Rev. D \textbf{102} (2020) no.2, 024005
[arXiv:2005.08828 [gr-qc]].

\bibitem{Nashed:2021ctg}
G.~G.~L.~Nashed and S.~Nojiri,
``Mimetic Euler-Heisenberg theory, charged solutions, and multihorizon black holes,''
Phys. Rev. D \textbf{104} (2021) no.4, 044043
[arXiv:2107.13550 [gr-qc]].

\bibitem{Karakasis:2022xzm}
T.~Karakasis, G.~Koutsoumbas, A.~Machattou and E.~Papantonopoulos,
``Magnetically charged Euler-Heisenberg black holes with scalar hair,''
Phys. Rev. D \textbf{106}, no.10, 104006 (2022)
[arXiv:2207.13146 [gr-qc]].

\bibitem{Barrientos:2016ubi}
J.~Barrientos, P.~A.~Gonz\'alez and Y.~V\'asquez,
``Four-dimensional black holes with scalar hair in nonlinear electrodynamics,''
Eur. Phys. J. C \textbf{76} (2016) no.12, 677
[arXiv:1603.05571 [hep-th]].

\bibitem{Gonzalez:2013aca}
P.~A.~Gonz\'alez, E.~Papantonopoulos, J.~Saavedra and Y.~V\'asquez,
``Four-Dimensional Asymptotically AdS Black Holes with Scalar Hair,''
JHEP \textbf{12}, 021 (2013)
[arXiv:1309.2161 [gr-qc]].

\bibitem{Olivares:2011xb}
M.~Olivares, J.~Saavedra, J.~R.~Villanueva and C.~Leiva,
``Motion of charged particles on the Reissner-Nordstr\"om (Anti)-de Sitter black holes,''
Mod. Phys. Lett. A \textbf{26} (2011), 2923-2950
[arXiv:1101.0748 [gr-qc]].

 \bibitem{Kuniyal:2015uta}
  R.~S.~Kuniyal, R.~Uniyal, H.~Nandan and K.~D.~Purohit,
  ``Null Geodesics in a Magnetically Charged Stringy Black Hole Spacetime,''
  arXiv:1509.05131 [gr-qc].

 \bibitem{Soroushfar:2016yea}
  S.~Soroushfar, R.~Saffari and E.~Sahami,
  ``Geodesic equations in the static and rotating dilaton black holes: Analytical solutions and applications,''
  Phys.\ Rev.\ D {\bf 94}, no. 2, 024010 (2016)
  [arXiv:1601.03143 [gr-qc]].

\bibitem{Gonzalez:2017kxt}
P.~A.~Gonz\'alez, M.~Olivares, E.~Papantonopoulos, J.~Saavedra and Y.~V\'asquez,
``Motion of magnetically charged particles in a magnetically charged stringy black hole spacetime,''
Phys. Rev. D \textbf{95}, no.10, 104052 (2017)
[arXiv:1703.04840 [gr-qc]].

\bibitem{Gonzalez:2015jna}
P.~A.~Gonzalez, M.~Olivares and Y.~Vasquez,
``Motion of particles on a Four-Dimensional Asymptotically AdS Black Hole with Scalar Hair,''
Eur. Phys. J. C \textbf{75} (2015) no.10, 464
[arXiv:1507.03610 [gr-qc]].

\bibitem{Gonzalez:2020kbv}
P.~A.~Gonz\'alez, M.~Olivares, E.~Papantonopoulos and Y.~V\'asquez,
``Timelike geodesics in three-dimensional rotating Ho$\check{\mathrm{r}}$ava AdS black hole,''
Phys. Rev. D \textbf{103}, no.8, 084037 (2021)
[arXiv:2008.00933 [gr-qc]].



\bibitem{Gonzalez:2019xfr}
P.~A.~Gonz\'alez, M.~Olivares, E.~Papantonopoulos and Y.~V\'asquez,
``Motion and trajectories of photons in a three-dimensional rotating Ho\v{r}ava-AdS black hole,''
Phys. Rev. D \textbf{101}, no.4, 044018 (2020)
[arXiv:1912.00946 [gr-qc]].



\bibitem{Gonzalez:2018lfs}
P.~A.~Gonz\'alez, M.~Olivares, E.~Papantonopoulos and Y.~V\'asquez,
``Motion and collision of particles in a rotating linear dilaton black hole,''
Phys. Rev. D \textbf{97}, no.6, 064034 (2018)
[arXiv:1802.01760 [gr-qc]].


\bibitem{Chatzifotis:2022ubq}
N.~Chatzifotis, N.~E.~Mavromatos and D.~P.~Theodosopoulos,
``Global Monopoles in the Extended Gauss-Bonnet Gravity,''
[arXiv:2212.09467 [gr-qc]].

\bibitem{Khodadi:2020jij}
M.~Khodadi, A.~Allahyari, S.~Vagnozzi and D.~F.~Mota,
``Black holes with scalar hair in light of the Event Horizon Telescope,''
JCAP \textbf{09} (2020), 026
[arXiv:2005.05992 [gr-qc]].






\end{thebibliography}
\end{document}